\documentclass{emulateapj}
\usepackage{graphicx}
\usepackage{amsfonts,amsmath,amssymb}
\usepackage{threeparttable}

\def\aj{{AJ}}		
\def\araa{{ARA\&A}}	
\def\apj{{ApJ}}		
\def\apjl{{ApJ}}	
\def\apjs{{ApJS}}	
\def\aap{{A\&A}}
\def\nat{{Nature}}	
\def\mnras{{MNRAS}}	
\def\icarus{{Icarus}}
\newcommand{\memb}{M_{\rm emb}}

\newcommand{\rgraze}{r_{\rm graze}}
\newcommand{\sigz}{\Sigma_z}
\newcommand{\au}{{\rm AU}}
\newcommand{\vesc}{v_{\rm esc}}
\newcommand{\vrel}{v_{\rm rel}}
\newcommand{\xmax}{x_{\rm max}}
\newcommand{\xmin}{x_{\rm min}}
\newcommand{\kep}{\emph{Kepler\ }}

\shorttitle{Compositions and Orbits}
\shortauthors{Dawson, Lee, \& Chiang}

\begin{document}
\slugcomment{Submitted to ApJ on December 13, 2015. Accepted by ApJ on March 22, 2016. In press.}
\title{Correlations between compositions and orbits \\established by the giant impact era of planet formation}

\author{Rebekah I. Dawson\altaffilmark{1,2}}
\author{Eve J. Lee}
\author{Eugene Chiang}
\affiliation{Department of Astronomy, University of California, Berkeley, 501 Campbell Hall \#3411, Berkeley CA 94720-3411}
\altaffiltext{1}{{\tt rdawson@psu.edu}}
\altaffiltext{2}{Department of Astronomy \& Astrophysics, The Pennsylvania State University; Center for Exoplanets and Habitable Worlds, The Pennsylvania State University}

\begin{abstract}
The giant impact phase of terrestrial planet formation 
establishes connections between super-Earths' orbital properties
(semimajor axis spacings, eccentricities, mutual inclinations) and
interior compositions (the presence or absence of gaseous envelopes).
Using $N$-body simulations and analytic arguments,
we show that spacings derive not only from eccentricities,
but also from inclinations.
Flatter systems attain tighter spacings, 
a consequence of an eccentricity equilibrium between
gravitational scatterings, which increase eccentricities, and mergers,
which damp them. Dynamical friction by residual disk gas plays a critical role in regulating mergers and
in damping inclinations and eccentricities. Systems with
moderate gas damping and high solid surface density
spawn gas-enveloped super-Earths with tight spacings,
small eccentricities, and small inclinations. Systems
in which super-Earths coagulate without as much ambient gas,
in disks with low solid surface density, produce rocky planets
with wider spacings, larger eccentricities, and larger 
mutual inclinations.
A combination of both populations
can reproduce the observed distributions of spacings, period ratios,
transiting planet multiplicities, and transit duration
ratios exhibited by \kep super-Earths.
The two populations, both
formed in situ, also help to explain observed
trends of eccentricity vs.~planet size, and bulk density vs.~method
of mass measurement (radial velocities vs.~transit timing variations). Simplifications made in this study --- including the limited timespan of the simulations, and the approximate
treatments of gas dynamical friction
and gas depletion history --- should be improved upon in future work to enable a detailed quantitative comparison to the observations.
\end{abstract}

\section{Introduction}

The \kep Mission discovered thousands of candidate ``super-Earths,'' planets between the size of Earth and Neptune (e.g., \citealt{Boru11a,Boru11b,Bata13,Burk14,Mull15}). 
They exhibit a wide range of bulk densities, anywhere
from $\sim$10 g/cm$^3$ to less than that of water (e.g., \citealt{Cart12,Wu13,Lope14,Weis14,Dres15,Wolf15}). Much of our knowledge of their orbital properties comes from the subset of systems each containing two or more transiting planets.
Statistical studies have found that systems of multiple super-Earth ``tranets'' (planets that transit; \citealt{Trem12})
have small eccentricities \citep{Moor11,Wu13,Hadd14,Vane15}
and small mutual inclinations \citep{Fang12b,Figu12,Fabr14} of a few percent
or less.

The orbital
spacings of tranets \citep{Liss11,Fang13,Fabr14,Liss14,Malh15,Stef15} have been a particular focus for theoretical studies of formation \citep{Hans13, Petr14,Malh15}, stability \citep{Pu15}, dynamical excitation \citep{Trem15}, and migration \citep{Lith12,Baty13,Deli14,Gold14,Hand14,Chat15,Deck15}. In this paper, we define the spacing
\begin{equation}
\Delta \equiv \frac{\delta a}{R_{\rm H}}
\end{equation}
as the semi-major axis difference $\delta a \equiv a_2 - a_1$ between
adjacent planets, normalized by their mutual Hill radius:
\begin{equation}
R_{\rm H} \equiv \frac{a_1+a_2}{2} \left(\frac{M_{{\rm p},1}+M_{{\rm p},2}}{3M_\star}\right)^{1/3},
\end{equation}
\noindent where $a$ is the semi-major axis, $M_{\rm p}$ is the planet mass, and $M_\star$ is the stellar mass.
The tranet spacing distribution peaks at $\Delta \sim 20$.
\citet{Fang13} found a similar spacing distribution for the underlying planets, assuming a single population of planetary systems drawn from independent distributions of mutual inclinations, spacings, multiplicities, and planet sizes. Planet spacings of $\Delta \sim 20$ --- shrinking
to $\Delta \sim 12$ for the highest multiplicity systems --- lie safely outside the empirical stability limit of $\Delta \sim 10$  \citep{Cham96,Yosh99,Zhou07,Smit09,Fang12,Liss14,Pu15}.
Studies of tranet multiplicity --- a property that depends on 
both spacing and mutual inclination --- have found a ``\kep dichotomy,'' i.e., a need for two populations to account for an apparent excess of single tranet systems
over multi-tranet systems \citep{Liss11,Joha12,Hans13,Ball14,Mori15}.

Here we explore how the orbital properties of super-Earths --- their
spacings, eccentricities, and inclinations --- originate from
the circumstances of their formation from primordial disks of solids
and gas. We will uncover correlations between these orbital properties
and planet compositions (gas-enveloped vs.~rocky), correlations
that are not due to observational bias and that are currently
neglected in \kep population studies (e.g., \citealt{Liss11,Fang12b,Figu12,Trem12,Fang13,Ball14,Fabr14,Stef15}). 
We work in the context of in-situ formation (e.g., \citealt{Hans12,Hans13,Chia13,Lee14,Daws15,Lee15,Lee16,Mori15}; see also \citealt{Inam15}), simulating the 
assembly of super-Earths starting from isolation masses.

We devote especial attention to the origin of super-Earth spacings.
The spacing distributions of planets and protoplanets have been explored
theoretically from several vantage points. One consideration is stability.
The timescale for orbit crossing is known empirically to increase
exponentially with the Hill spacing $\Delta$ \citep{Yosh99,Zhou07,Pu15}.
The spacing must be wide enough for the orbit crossing time to exceed the
system's age. \citet{Trem15} presented another point of view from
statistical mechanics. Other perspectives derive from considerations
of planet formation. For isolation masses (a.k.a.~oligarchs) accreting small
bodies in their feeding zones, a spacing equilibrium is achieved between
mass growth, which decreases the separation in Hill radii, and orbital
repulsion driven by gravitational (a.k.a.~viscous) stirring
\citep{Koku95,Koku98}.
In the particle-in-a-box approximation for planet formation by coagulation
of small bodies (e.g.,
\citealt{Safr69,Gree90,Ohts92,Ford01,Gold04,Ida13,Petr14,Joha12,Morr15}),
random velocities are 
limited to the surface escape velocity $\vesc$
from the most massive body.
\citet{Schl14} applied this principle to estimate the minimum disk masses required to form super-Earths in situ.
A planet that excites the random velocities of surrounding bodies to
the maximum value of $\vesc$ has a feeding zone of full width
\begin{equation}
\Delta = \frac{2\vesc}{n R_{\rm H}} \simeq 45\left(\frac{P}{\rm yr}\right)^{1/3} \left(\frac{\rho}{{\rm g\, cm}^{-3}}\right)^{1/6},
\end{equation}
\noindent where $n$ is the planet's orbital angular frequency (a.k.a.~mean
motion), $P = 2\pi/n$ is the orbital period, and $\rho$ is the planet's bulk
density. This feeding zone width also corresponds to the expected maximum spacing between nascent planets. We will take a related approach by finding the spacing that allows for an
``eccentricity equilibrium'': one that balances gravitational scatterings
between protoplanets, which
excite eccentricities, with mergers, which damp them.

Our paper, which explores how the orbital spacings of super-Earths
are fossil records of their formation in situ, and how orbital
properties in general are correlated with planet composition, 
is organized as follows.
In Section 2, we describe the setup for the $N$-body simulations 
that are the basis for all subsequent sections.
In Section 3, we show how spacings, eccentricities, and inclinations
evolve with time for a few illustrative, gas-free simulations.
In particular, we demonstrate how the evolution is sensitive
to initial inclinations --- an input parameter that, to our knowledge,
is not often highlighted as a controlling parameter in coagulation
calculations. We offer order-of-magnitude scalings to understand
these results, explaining the dependence of spacings on inclinations
in terms of an eccentricity
equilibrium that balances scatterings with mergers.
Section 4 gives a more complete overview of our gas-free simulations,
detailing how final spacings depend on various initial conditions.
Section 5 brings dynamical friction by disk gas into the mix; we investigate
how gas can establish those initial conditions that were assumed
in preceding sections. Including gas also enables us to connect planet
compositions --- whether or not planets have volumetrically significant
gas envelopes --- with orbital properties; this is the focus of Section 6.
Comparisons with observations are made in Section 7; there we
will find that we need a mixture of dynamically hot 
and dynamically cold populations to more faithfully reproduce
\kep data. We present our conclusions in Section 8.

\section{Initial conditions for $N$-body simulations}

To assess how orbital spacings depend on conditions
during the late stages of planet formation,
we simulate the growth of isolation mass embryos to
super-Earths via collisions (a.k.a.~giant impacts). 
Building on work by, e.g., \citet{Cham96}, \citet{Koku98,Koku02}, 
\citet{Hans12,Hans13}, and \citet{Daws15},
we perform $N$-body integrations each lasting 27 Myr 
using the hybrid symplectic integrator of
{\tt mercury6} \citep{Cham99}.
The time step is 0.5 days and a close encounter
distance (which triggers a switch from the symplectic
integrator to the Burlisch-Stoer integrator)
is 1 $R_{\rm H}$.
We run several thousands of simulations, grouped into
ensembles and summarized in Table \ref{tab:ens}. The default number of simulations per ensemble is 80.

Our simulations begin with embryos each having an
isolation mass $\memb$:
\begin{eqnarray}
\label{eqn:memb}
\memb =
0.16 M_{\oplus} \left(\frac{\Delta_0}{10}\right)^{3/2}\left(\frac{\Sigma_{z,1}}{10\, {\rm g}/{\rm cm}^2}\right)^{3/2} \nonumber \\
\times \left( \frac{a}{\au}\right)^{3(2+\alpha)/2} \left( \frac{M_\star}{M_\odot} \right)^{-1/2} 
\end{eqnarray}
\noindent where $\sigz = \Sigma_{z,1} (a/{\rm AU})^{\alpha}$. We explore a large range of $\Sigma_{z,1} = 3$--400 g/cm$^2$. By default, we use an initial spacing $\Delta_0=10$, similar to \citet{Hans12,Hans13}, and based on the \citet{Koku98,Koku02} simulations of embryos formed via the accretion of small bodies.  We list the range of the initial number of embryos per simulation, $N_{\rm emb,0}$, and the final number of planets, $N_{\rm p,final}$, in Table 1. Some of our simulations include damping by
residual disk gas; these are detailed
in Section \ref{sec:damp}.
We assign planets constant bulk densities $\rho$ that
define their physical cross sections. When two embryos
touch, we assume perfect accretion with no
fragmentation.  

Initial eccentricities and inclinations are specified in Table \ref{tab:ens}. Depending on the ensemble, the magnitude of the initial eccentricity $e_0$ is set either to 0; to a constant fraction (specified
in the Table) of
${h \equiv [(M_{\rm emb,1} + M_{\rm emb,2})/(3M_\star)]^{1/3}}$, 
where the subscripts ``1'' and ``2'' refer to adjacent inner and
outer bodies;
or to a constant fraction of the orbital separation
${\delta a / a \equiv (a_2-a_1)/[2(a_2+a_1)]}$. 
Flat ensembles ({\tt Ef, Eci0}) are strictly 2D, beginning with and maintaining $i=0$. For other ensembles, the magnitude of the initial inclination $i_0$ is drawn either from a uniform distribution between 0--0.1$^\circ$ or 0--0.001$^\circ$; set to a constant value
$e_0/\sqrt{2}$ (equipartition); or set to $0.01h$. 
The initial mean anomaly, argument of periapse, and longitude of ascending node are drawn randomly from a uniform distribution spanning 0--2$\pi$. In the remainder of this work, we compute the reported $i$ of each planet relative to the initial $i=0$ plane. 

Following \citet{Hans12,Hans13}, the inner edge of the disk of embryos is drawn randomly from a uniform distribution spanning 0.04--0.06 AU. Other embryos are spaced $\Delta_0$ away from each other, up to a maximum $a$ of 1 AU. Embryos are initially placed in order of increasing $a$. The semi-major axis $a_j$ and mass $M_{{\rm emb},j}$ of the $j$th embryo are computed from the mutual Hill radius of the previous embryo, i.e., $R_{\rm H} = a_{j-1} \left(\frac{2M_{{\rm emb},j-1}}{3M_\star}\right)^{1/3}$. As a result, the actual initial orbital spacings differ slightly from the values of $\Delta_0$ listed in Table 1. The surface density normalization $\Sigma_{z,1}$
for each simulation in the ensemble is drawn from a uniform log distribution spanning the range specified in Table 1.
The surface density slope $\alpha=-3/2$ except for ensemble
{\tt Eh$\alpha$-2}, which
uses $\alpha = -2$, as noted in Table 1.

\begin{table*}
\caption{Ensembles of Simulations \label{tab:ens}}
\centering
\begin{threeparttable}
\begin{tabular}{@{}lccccccccc@{}}
\hline
Name & $e_0$ & $i_0$ & $\Delta_0$ & $\Sigma_{z,1}$ & $N_{\rm emb,0}$ & $N_{\rm p,final}$ &Notes\\ 
&  & (rad)&$(R_{\rm H})$ & (g cm$^{-2}$)  \\
\hline
{\tt Ef}  &0&0&10&33--90&22--35&7--13 \\
{\tt E}  
&0&$0.1^{\circ}$&10&33--90&22--35&4--11&a
\\
{\tt Eh}  &$\sqrt{2}h/\sqrt{3}$&$e_0/\sqrt{2}$&10&3--403 &12--128&2--22& b,c\\ 
{\tt Eh$\alpha$-2}  &$\sqrt{2}h/\sqrt{3}$&$e_0/\sqrt{2}$&10&33--55 &19--25&6--8& c,d\\ 
{\tt Eh$\rho$}  &$\sqrt{2}h/\sqrt{3}$&$e_0/\sqrt{2}$&10&116&20--23&4--7&c,e\\  
{\tt Ec}  &$\delta a/(2a)$&$e_0/\sqrt{2}$&10&33--90&22--35& 4--10
&f\\
{\tt Eci0}  &$\delta a/(2a)$&0&10&33--90&22--35&5-11&f\\
{\tt Eci0.1}  &$\delta a/(2a)$&$0.1^{\circ}$&10&33--90&22--35& 4--8& a,f\\
{\tt Eci0.001}  &$\delta a/(2a)$&$0.001^{\circ}$&10&33--90&22--35&4-12&a,f\\
{\tt Ece0.3}  &$0.3 \delta a/a$&$e_0/\sqrt{2}$&10&33--90&22--35&4--9&f\\
{\tt Ece0.1}  &$0.1 \delta a/a$&$e_0/\sqrt{2}$&10&33--90&22--35&4--9&f\\
{\tt Ece0.3i0.1}  &$0.3 \delta a/a$&$0.1^\circ$&10&33--90&22--35&5--8&a,f\\
{\tt Ece0.1i0.1}  &$0.1 \delta a/a$&$0.1^\circ$&10&33--90&22--35&5--10&a,f\\
{\tt E$\Delta$3}  &0&$0.1^{\circ}$&3&33--90&129--240&5--9&a\\
{\tt E$\Delta$5}  &0&$0.1^{\circ}$&5&33--90&62--111&5--8&a\\
{\tt E$\Delta$7.5}  &0&$0.1^{\circ}$&7.5&33--90&34--54&4--8&a\\
{\tt E$\Delta$9}  &0&$0.1^{\circ}$&9&33--90&27--43&3--8&a\\
{\tt E$\Delta$11.5}  &0&$0.1^{\circ}$&11.5&33--90&19--33&4--12&a\\
{\tt E$\Delta$13}  &0&$0.1^{\circ}$&3&33--90&17--28&3--14&a\\
{\tt Ed10$^0$}  & $h$ & $h/100$ &3&38--105&123--220&2--9&c,g\\
{\tt Ed10$^1$}  & $h$ & $h/100$ &3&38--105&123--220&3--9&c,g\\
{\tt Ed10$^2$}  & $h$ & $h/100$ &3&38--105&123--220&3--18&c,g\\
{\tt Ed10$^3$}  & $h$ & $h/100$ &3&38--105&123--220&2--9&c,g\\
{\tt Ed10$^4$} & $h$ & $h/100$ &3&38--105&123--220&2--9&c,g\\
\hline
\end{tabular}
\begin{tablenotes}
\item[a]{Inclinations drawn randomly from a uniform distribution between 0 and $i_0$.}
\item[b]{Contains 500 simulations. The default is 80.}
\item[c]{$h \equiv [(M_{{\rm p},1}+M_{{\rm p},2})/(3M_\ast)]^{1/3}$ is the Hill parameter.}
\item[d]{Surface density power-law slope is $\alpha=-2$ instead of the default $\alpha=-3/2$.}
\item[e]{The planet bulk density $\rho$ is drawn from a uniform log distribution from
0.02--14 g cm$^{-3}$.
The default is a fixed $\rho = 1$ g cm$^{-3}$.}
\item[f]{$\delta a \equiv a_2 - a_1$ is the semi-major axis difference between neighboring embryos.}
\item[g]{Includes gas damping. See Section 5.}

\end{tablenotes}
\end{threeparttable}
\end{table*}

\section{Growth and equilibration of eccentricities and spacings} \label{sec:analytic}

\begin{figure}
\begin{center}
\includegraphics{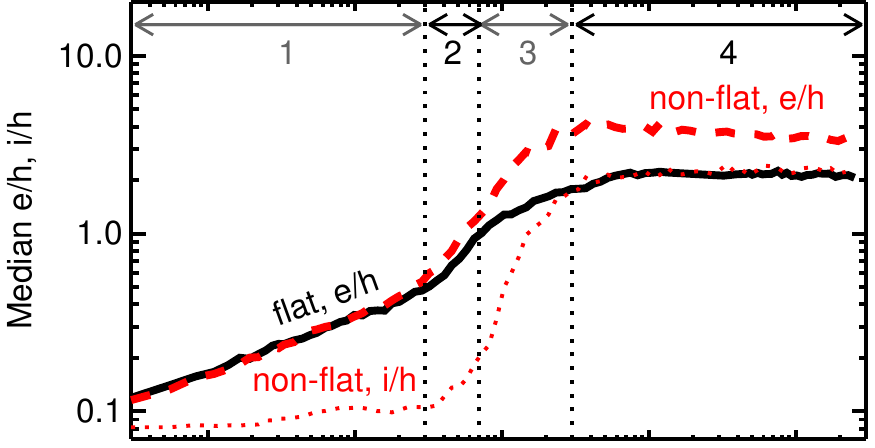}
\includegraphics{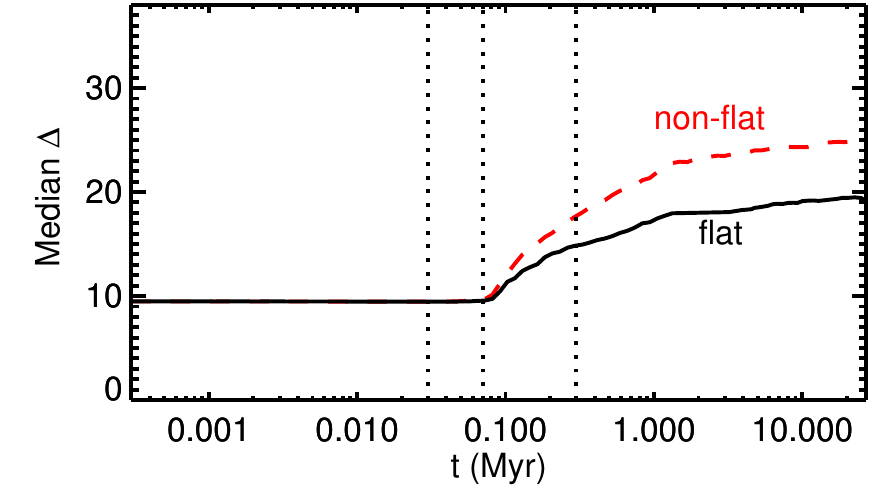}
\includegraphics{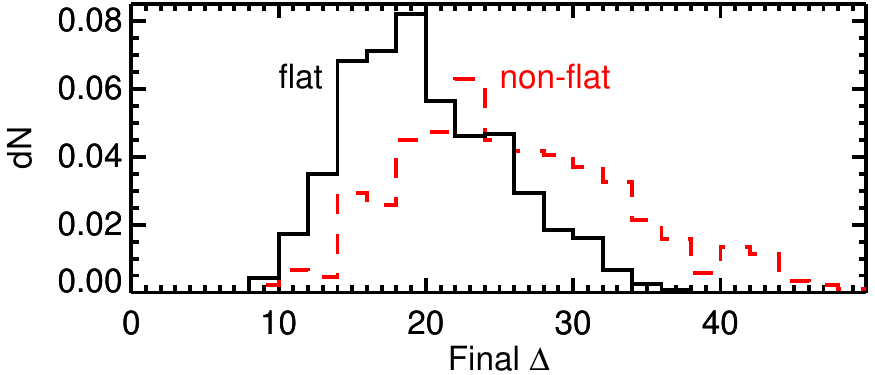}
\caption{
 \label{fig:avgevo} 
 Inclinations matter for orbital spacings. Top: Evolution of $e$ (thick curves) and $i$ (thin curve), averaged over all planets in an ensemble. Averages from the non-flat ensemble {\tt E} are plotted as red dashed curves and from flat ensemble {\tt Ef} as a black solid curve; $i$ does not appear on the plot for {\tt Ef} because that ensemble is completely 2D $(i=0)$. For both ensembles, the eccentricity undergoes four stages of evolution, as demarcated by vertical dotted lines. Middle: Same for Hill spacings $\Delta$ for adjacent pairs of planets. Planet pairs in the flat ensemble (black solid) end up with tighter spacings. Bottom: Distributions of final $\Delta$.
}
\end{center}
\end{figure}

\begin{figure*}
\begin{center}
\includegraphics[width=\textwidth]{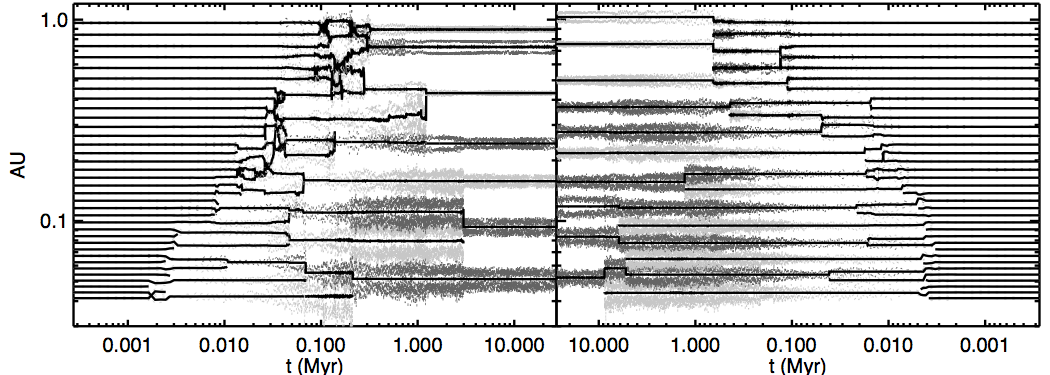}
\caption{
\label{fig:evo} Evolution of $a$ (black) and $a (1\pm e)$ (gray). Left: $\Sigma_{z,1} = 55$ g/cm$^2$ simulation from {\tt Eh}. The final planets have $ 1.5 < M_p < 5 M_\oplus$. Right: Identical simulation with initial $i=0$. The planets end up more tightly spaced (in both $a$ and $\Delta$). 
}
\end{center}
\end{figure*}

During the giant impact stage of planet formation, planets scatter and merge, establishing the system's orbital spacings and eccentricities. Through mutual gravitational interactions, planets convert Keplerian shear into random 
velocity, leading to growth in $e$ and $i$. Mergers (inelastic collisions) counter the growth of random velocities and stabilize the system by widening the spacings between bodies. 
Figure \ref{fig:avgevo} (row 1) compares the average eccentricity growth from two ensembles of simulations ({\tt E, Ef}), two individual members of which are shown in Figure \ref{fig:evo}. Planets begin as moon-to-Mars mass embryos on circular orbits and grow in mass, separation, and eccentricity. Fig.~\ref{fig:avgevo} shows four stages for eccentricity growth: the initial growth at low $e$ (stage 1), a faster growth as $e$ approaches orbit crossing (stage 2), growth during orbit crossing (stage 3), and an eccentricity equilibrium (stage 4). For the non-flat ensemble, the average inclination is also plotted in Figure \ref{fig:avgevo}. The second row of Figure \ref{fig:avgevo} shows the evolution of $\Delta$, the spacing in mutual Hill radii, which is flat throughout the first two eccentricity growth stages when mergers do not yet occur.

 In this section we use order-of-magnitude
calculations to understand how
eccentricities, inclinations,
and orbital spacings are connected.
We focus exclusively on the latter phases
of the evolution --- stages 3 and 4 ---
when mergers occur and spacings $\Delta$
evolve from their initial values.

\subsection{Eccentricity growth rate}
\label{sec:grow}

 We derive an order-of-magnitude formula
for eccentricity growth for use in later
subsections when we consider the specifics
of stages 3 and 4.
Consider widely-spaced,
shear-dominated pairs for which the relative
velocities are
\begin{equation}
\vrel \sim v_{\rm H} \Delta,
\end{equation}
\noindent where $v_{\rm H}$ is the mutual Hill velocity.
The relative velocity changes impulsively every conjunction.
An encounter between
two planets each of mass $M_{\rm p}$
at impact parameter $x$ produces an acceleration $$\frac{GM_{\rm p}}{x^2} =\frac{3 n^2 R_{\rm H}^3}{2 x^2}$$ over a time interval $2 x/\vrel$, changing $\vrel$ by
\begin{equation}
\label{eqn:del}
\delta \vrel = \frac{3 n^2 R_{\rm H}^3}{ x \vrel}.
\end{equation}
It is assumed that $\delta \vrel$
is randomly directed;
over many encounters, $\vrel$ random walks.
For a given $x$, the expected number of synodic periods $N_{\rm synodic}$
(i.e., the number of conjunctions or random walk steps) required to 
change $\vrel$ by a fractional amount
$f$ is given by
\begin{equation}
\label{eqn:nsyn}
\frac{1}{N_{\rm synodic}} = \left(\frac{\delta \vrel}{f \vrel}\right)^2 =\frac{9 R_{\rm H}^2}{f^2 x^2 \Delta^4}.
\end{equation}

\begin{figure*}
\begin{center}
\includegraphics[width=.45\textwidth]{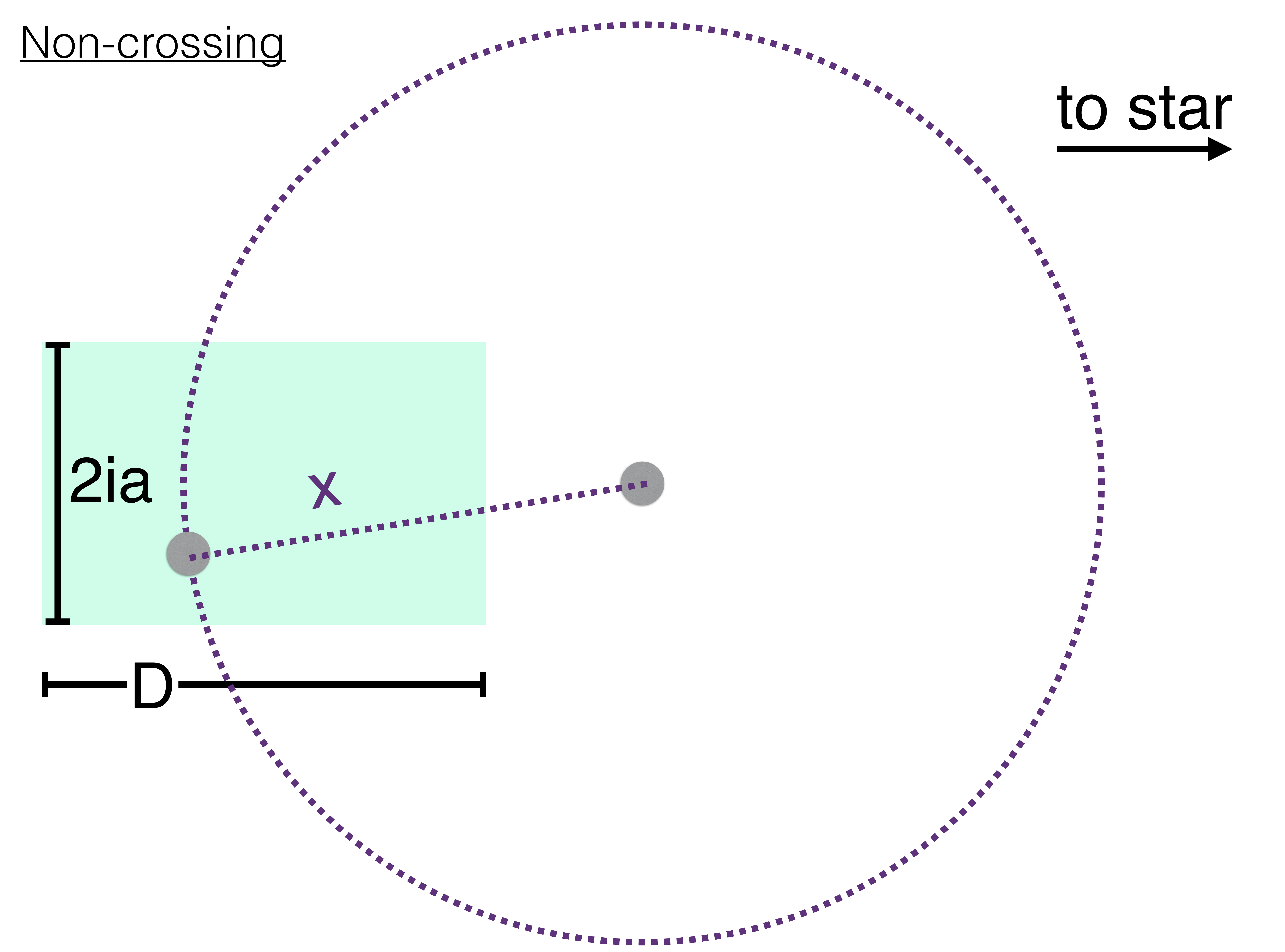} \includegraphics[width=.45\textwidth]{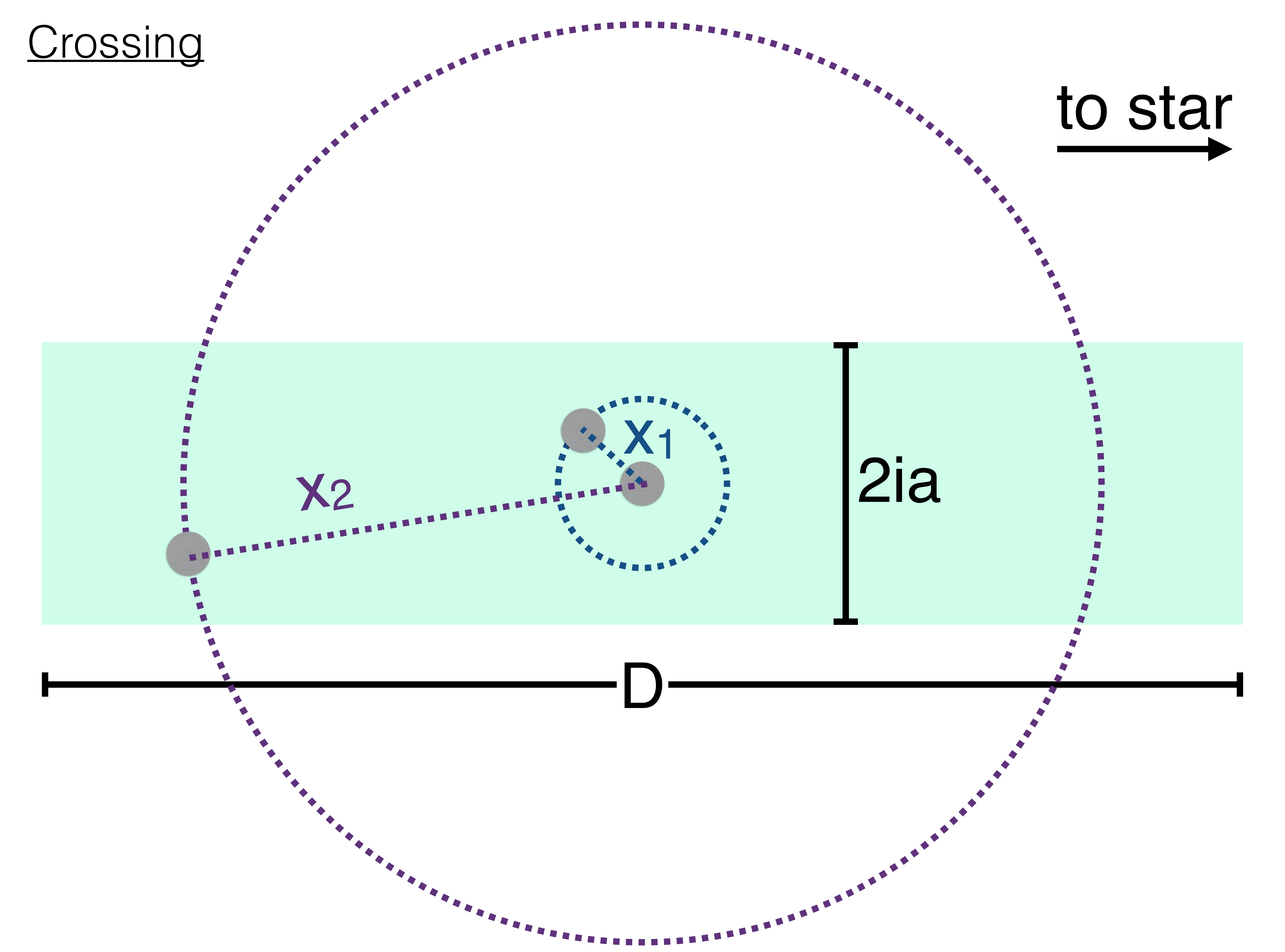}
\caption{
 \label{fig:icartoon} The probability of an encounter with impact parameter $x$ between two bodies depends on the value of $x$ relative to the scale height $ia$. The rectangle represents the area of possible encounter space for the pair of bodies. Left: Non-crossing orbits. The probability of an encounter within $dx$ of $x$ is 
 $\frac{2 x \sin^{-1} \left[ia/x\right] dx}{ ia D} \sim \frac{dx}{D}$. 
 Right: Crossing orbits. For $x_1 < ia$, the probability of an encounter within $dx$ of $x_1$ is $\frac{2\pi x_1 dx}{2ia D} \sim \frac{x_1 dx}{ia D}$, where $D$ is the range of 
 horizontal separations.  For $x_2>ia$, the probability of an encounter within $dx$ of $x_2$ 
 is $\frac{4 x_2 \sin^{-1} \left[ia/(2x_2)\right] dx}{2 ia D} \sim \frac{2 dx}{D}$. 
 Hence the encounter probability depends on 
 $i$ for $x<ia$ but not for $x>ia$.
}
\end{center}
\end{figure*}

Eqn.~\ref{eqn:nsyn} gives a rate of viscous stirring, which we now average
over the geometry of possible impact parameters $x$ from $\xmin$ to $\xmax$. The probability of an encounter with impact parameter $x$ depends on the value of $x$ relative to the ``scale height'' $ia$, as depicted in Figure \ref{fig:icartoon}.
The scale height for a pair of bodies reflects the range of mutual inclinations resulting from their inclinations $\sim i$ with respect to the initial plane and their range of nodal orientations. Let $D$ be the range of horizontal separations, determined by $\Delta$ and the eccentricity vectors. For $x > ia$, the dynamics is essentially 2D (i.e., in a common orbital plane) and independent of the inclinations.
If orbits are crossing (Figure \ref{fig:icartoon}, right panel),
the probability of a conjunction occurring within the interval $x$ and $x+dx$ for $x > ia$
is
$$ p(x) dx = \frac{4 x \sin^{-1} \left(ia/x\right) dx }{2 ia D} \sim \frac{2 dx}{D},\,\,\,\, x>ia, \,{\rm crossing}.$$
For non-crossing orbits (Figure \ref{fig:icartoon}, left panel), this 2D probability is reduced by a factor of 2.
For $x < ia$ (the 3D case), the corresponding probability is
$$ p(x) dx = \frac{2\pi x \,dx}{2ia D}  \sim \frac{\pi x\,dx}{ia D} ,\,\,\,\, x< ia, $$ reduced by a factor of $\pi x/(2ia)$ relative to the 2D case and dependent on $i$. 

Weighting Eqn.~\ref{eqn:nsyn} by these probabilities and integrating from $\xmin$ to $\xmax$, we find
\begin{eqnarray}
\label{eqn:pstir}
\Bigg \langle\frac{1}{N_{\rm synodic}}\Bigg \rangle_{\rm stir} &=& \frac{9 R_{\rm H}^2}{f^2 D\Delta^4}\nonumber \\
&&\times \Bigg ( \int_{\max (\rgraze,\xmin)}^{\max [\rgraze,\xmin,\min(ia,\xmax)]} \frac{\pi dx}{ia x}\nonumber\\
&&+\int^{\xmax}_{\max [\rgraze,\xmin,\min(ia,\xmax)]} \frac{2 dx}{x^2} \Bigg) \nonumber\\\
&=&\frac{18 R_{\rm H}^2}{f^2 D\Delta^4} \nonumber \\ 
&& \times \Bigg ( \frac{\pi}{2 ia} \ln\left[ \frac{\max \left[\rgraze,\xmin,\min(ia,\xmax)\right]}{\max (\rgraze,\xmin)}\right] \nonumber \\
&&+  \frac{1}{\max \left[\rgraze,\xmin,\min(ia,\xmax)\right]}-\frac{1}{\xmax} \Bigg), \nonumber\\
\end{eqnarray}
\noindent where the impact parameter for two bodies of radius $R_{\rm p}$ to undergo
a grazing collision is
\begin{eqnarray}
\label{eqn:rgraze}
\rgraze &=& 2 R_{\rm p} \left[1+(\vesc/\vrel)^2\right]^{1/2} \nonumber \\
&=& \frac{12 R_{\rm H}}{\Delta} \left(\frac{v_{\rm H}}{2 \vesc}\right) \left[1+4\left(\frac{\Delta}{2 \vesc/v_{\rm H}}\right)^2\right]^{1/2}. \nonumber \\
\end{eqnarray}
\noindent The $\max (\rgraze,\xmin)$ lower limit of the first integral applies because when $x<\rgraze$ at conjunction, a merger occurs instead of a scattering. The $\max [\rgraze,\xmin,\min(ia,\xmax)]$ upper limit of the first integral 
ensures that only the encounters with $x < ia$ are taken into account. For $\xmin > ia$, the first integral vanishes.
Conversely, the second integral, which 
takes into account encounters with $x > ia$,
vanishes if $\xmax < ia$.

Eqn.~\ref{eqn:pstir} applies to the growth of random velocity --- i.e., both $e$ and $i$ --- but in the remainder of this section we will apply it to eccentricity growth.  Most encounter geometries either predominantly excite $e$ ($x>ia$)
or excite both $e$ and $i$ by comparable amounts ($x<ia$).

\subsection{Stage 3: Eccentricity growth post-orbit crossing}

In stage 3, orbits cross and bodies merge (Fig.~1: $\Delta$ is increasing in row 2 during this stage). 
Eccentricity excitation by scatterings is tempered
by eccentricity damping by mergers.
Eccentricity damping follows from conservation of momentum:
an inelastic head-on collision between two equal-mass
bodies with relative velocity $v_{\rm rel}$ produces
a single body with velocity $v_{\rm rel}/2$.
For $e \propto \vrel$, this argument yields
$\delta e_{\rm merger} \approx -0.5 e$. 
Empirically, we determine from our simulations (Sections 2 and 4)
that $\delta e_{\rm merger} \approx -0.4e$ is a more accurate
rule of thumb. (See \citet{Mats15} for a detailed study of eccentricity damping via giant impacts.) The
takeaway
point is that each merger
lowers the eccentricity of the colliding pair
by a factor of order unity. Thus
the number of conjunctions (i.e., synodic periods) required
to reduce the median eccentricity by order unity is the inverse
of the probability of a merger per conjunction, i.e., the inverse
of the probability that $x \leq \rgraze$ in a given encounter:
\begin{eqnarray}
\label{eqn:pmerge}
\Bigg \langle\frac{1}{N_{\rm synodic}}\Bigg \rangle_{\rm damp} &\sim& \frac{\pi \rgraze^2}{D ia}, \,\,\,\, ia > \rgraze \nonumber \\
&\sim& \frac{2 \rgraze}{D}, \,\,\,\, ia < \rgraze. \nonumber \\
\end{eqnarray}
\noindent The two cases correspond to 3D and 2D encounters,
respectively.

We compare the above damping rate to the stirring rate (Eqn. \ref{eqn:pstir}),
evaluated for $\xmin < \rgraze$ and $\xmax > ia$:
\begin{eqnarray}
\label{eqn:pstir3}
\Bigg \langle\frac{1}{N_{\rm synodic}}\Bigg \rangle_{\rm stir,3}\sim \frac{72 R_{\rm H}^2}{D\Delta^4} \times \nonumber \\
\Bigg \{ \frac{\pi}{2ia} \ln\left[ \frac{\max \left(ia,\rgraze\right)}{\rgraze}\right]
+  \frac{1}{\max \left(ia,\rgraze\right)}-\frac{1}{\xmax} \Bigg \}, \nonumber\\
\end{eqnarray}
\noindent where we have set $f=0.5$ because we are interested
in order-unity changes to the eccentricity.

As $\xmax \rightarrow \infty$, the ratio of the stirring rate to the damping rate is
\begin{eqnarray}
\label{eqn:equal}
\frac{\left \langle\frac{1}{N_{\rm synodic}} \right \rangle_{\rm stir}}{ \left \langle\frac{1}{N_{\rm synodic}} \right \rangle_{\rm damp}} & \nonumber \\
\sim & \frac{72 R_{\rm H}^2}{\pi \Delta^4\rgraze^2} \left[\frac{\pi}{2}\ln\left(\frac{ia}{\rgraze}\right) + 1 \right]&, \,\,ia > \rgraze \nonumber \\
\sim & \frac{36 R_{\rm H}^2}{ \Delta^4\rgraze^2}&, \,\,ia < \rgraze. \nonumber \\
\end{eqnarray}
\noindent Note how $D$ conveniently divides out. 
 In stage 3, the ratio given
by Eqn.~\ref{eqn:equal} is greater than unity: stirring exceeds damping for small
$\Delta$, and eccentricities and inclinations rise.
At a given $\Delta$, the stirring-to-damping ratio is larger --- by a logarithm --- for $ia > \rgraze$ than for $ia < \rgraze$.
This is consistent with Fig.~1 (row 1), which shows that
eccentricity growth is somewhat faster
in non-flat systems as compared to flat systems during stage 3.

\subsection{Stage 4: Eccentricity equilibrium}
\label{sec:equil}

An eccentricity equilibrium can be achieved when the rate of eccentricity growth from scatterings matches the rate of eccentricity damping from mergers. This eccentricity equilibrium
is evident in stage 4 of Fig.~1 (row 1). It is  
distinct from the spacing equilibrium discussed
by \citet{Koku95,Koku98}, whereby orbital repulsion
between big bodies (driven by scattering and dynamical friction) keeps pace
with the expansion of their Hill radii caused by
accretion of small bodies.

Setting the stirring-to-damping ratio\footnote{This ratio was derived under the assumption that $\xmax > ia$. We do not treat
the case $\xmax < ia$.} in Eqn.~\ref{eqn:equal} to unity gives the value of $\Delta$ necessary to achieve an eccentricity equilibrium:
\begin{eqnarray}
\left(\frac{\Delta}{2 \vesc/v_{\rm H}}\right) \left[1+4\left(\frac{\Delta}{2 \vesc/v_{\rm H}}\right)^2\right]^{1/2} =\nonumber\\ 
\frac{1}{2}   \sqrt{ \ln\left( \frac{ia}{\rgraze}\right) \nonumber  + \frac{2}{\pi} }, \,\,\,\,ia > \rgraze \nonumber \\
\frac{1}{2} , \,\,\,\,ia < \rgraze \nonumber \\
\end{eqnarray}
or
\begin{eqnarray}
\label{eqn:deltaover}
\frac{\Delta}{2 \vesc/v_{\rm H}} = 0.5 \sqrt{\sqrt{ \ln\left( \frac{ia}{\rgraze}\right) + 0.89}-0.5},\,\,\,\, ia> \rgraze \nonumber \\
\frac{\Delta}{2 \vesc/v_{\rm H}} = 0.39,\,\,\,\, ia< \rgraze \nonumber \,.\\
\end{eqnarray}
Eqn.~\ref{eqn:deltaover}
can be re-cast as:
\begin{eqnarray}
\label{eqn:delta}
 \Delta = & 22 \left(\frac{P}{\rm yr}\right)^{1/3} \left(\frac{\rho}{{\rm g}/{\rm cm}^{3}}\right)^{1/6} \nonumber \\ 
 & \times \sqrt{\sqrt{\ln\left( \frac{ia}{\rgraze}\right) + 0.89}-0.5}, \,\,\,\,ia > \rgraze, \nonumber \\
  \Delta = & 17 \left(\frac{P}{\rm yr}\right)^{1/3} \left(\frac{\rho}{{\rm g} /{\rm cm}^{3}}\right)^{1/6} , \,\,\,\,ia < \rgraze \,.
 \end{eqnarray}

At the risk of over-interpreting our order-of-magnitude
calculations, we infer from
Eqns.~\ref{eqn:deltaover} and \ref{eqn:delta} that: (a) the
spacing that gives an eccentricity equilibrium (hereafter
the ``equilibrium spacing'') does not
depend explicitly on planet mass, embryo surface density,
or (for a given orbital period) stellar mass;
(b) the equilibrium spacing does depend on planet bulk density
because lower $\rho$ corresponds, at fixed mass, to
larger $R_{\rm p}$ and therefore larger $\rgraze$
(Eqn.~\ref{eqn:rgraze}); larger merger rates must be balanced
by larger stirring rates which are obtained for smaller $\Delta$
(compare Eqns.~\ref{eqn:pmerge} and \ref{eqn:pstir3}); (c) for
practically 
all values of $i$, 
the equilibrium spacing
is less than the standard quoted value of
$2v_{\rm esc}/v_{\rm H}$ (Section 1), in agreement
with our numerical simulations (Section 4)
and observations (e.g., \citealt{Fang12,Pu15}).

Of especial note is (d) higher 
$i$'s lead to wider spacings. While
stirring and merger rates are each reduced by
increasing 
$i$ --- because
the volume of space in which bodies
interact $\propto ia$ ---
the stirring rate is reduced less severely,
because encounters occurring at
the largest impact parameters,
at $x > ia$, unfold independently of $i$.
For example, for $ia = 10 \rgraze$, Eqn.~\ref{eqn:delta} yields a spacing about 50\% wider than
in a flat ($ia < r_{\rm graze}$) system.

 Our considerations of orbital spacings are qualitatively
similar to those of \citet{Pu15}. Both of our studies
recognize that increasing mutual inclinations in packed
multi-planet systems demands larger orbital spacings
to maintain dynamical stability (cf.~our Eqn.~\ref{eqn:delta}
with their equation 14). Since their study is numerical,
it accounts for effects not captured by the crude
arguments we have made in this Section \ref{sec:analytic}.
In particular, our treatment above assumes that each
conjunction between neighboring planets is a step in a random
walk; this assumption may fail at large orbital spacings where
the dynamics is less chaotic. Hopefully this shortcoming
does not compromise our goal to understand,
if only qualitatively, how
orbital spacings are determined
during the formative stages of
planetary systems, when spacings are comparatively small.
We offer Eqn.~\ref{eqn:delta} as a plausibility argument
that can explain some of the results of our numerical
simulations. These simulations are the focus for the remainder
of our paper.

\section{Dependence of spacings
on initial (post-damping) conditions}

Here we use simulations to explore how orbital spacings depend on initial conditions in the purely $N$-body regime, i.e., after damping from residual gas/planetesimals has ceased. In this section, we assign a range of ad hoc initial conditions. In Section 5, we will explore how gas damping might establish such initial conditions.

\subsection{Systems flatter during orbit crossing end up with tighter spacings}

In Section 3.5, we demonstrated with scaling arguments that the spacing necessary to achieve an eccentricity equilibrium via stirring and mergers depends on the 
planets' inclinations. 
We found that the final spacings of flat systems $(i=0)$ are tighter than those of systems where $i$ is free to grow (Fig.~1). Here we compare several additional ensembles that all have $i$ free to grow but have different 
$i$'s 
during the orbit crossing stage, when 
$i$ 
starts to affect the stirring rate (i.e., eccentricity growth stage 3, Section 3.4).

\begin{figure}
\begin{center}
\includegraphics{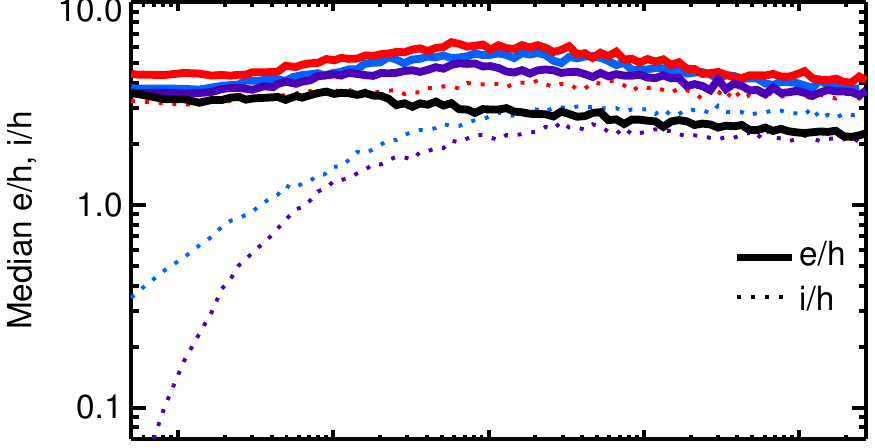}
\includegraphics{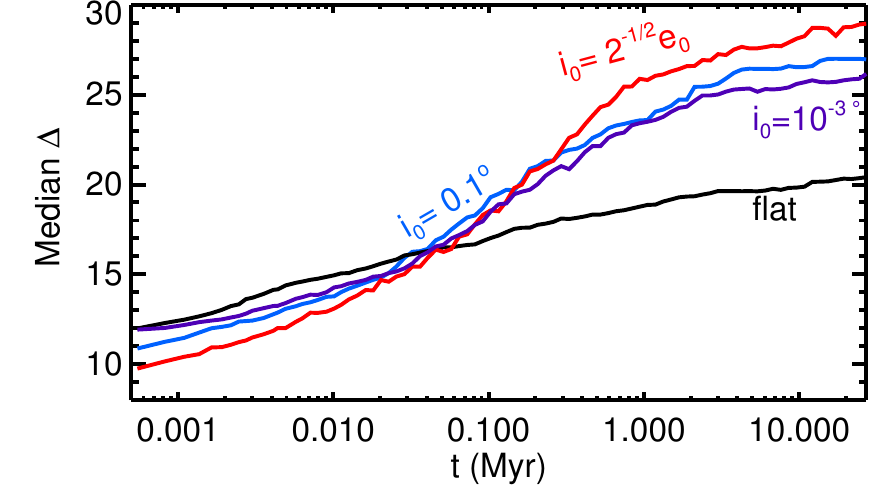}
\caption{
 \label{fig:icross}  
 Inclinations
 when orbits start to cross affect final spacings. Each of the ensembles (red, blue, purple, black) begins with an identical distribution of initial conditions excepting
inclinations.
Top: Evolution of $e$ (solid) and $i$ (dotted), averaged over all planets in an ensemble. The eccentricity evolution is similar across ensembles except the $i=0$ ensemble (black). 
Bottom: The 
smaller the initial $i$,
the tighter the final spacing. Each simulation begins with $\Delta_0=10$ at $t=0$.
}
\end{center}
\end{figure}

Each ensemble of simulations begins with planets with eccentricities large enough to cross:
$$e_0 = 0.5 \,\delta a/a,$$
where $\delta a$ is the difference in semi-major axes between adjacent bodies. The four ensembles differ in the magnitudes of their initial 
inclinations: $i_0=0$ ({\tt Eci0}, black curve in Fig.~\ref{fig:icross}); $i_0=0.001^\circ$
({\tt Eci0.001}, purple), $i_0 = 0.1^\circ$ ({\tt Eci0.1}, blue), and $i_0 = e_0/\sqrt{2}$ ({\tt Ec}, red).

We plot the evolution of $e$, $i$, and $\Delta$ in Fig.~\ref{fig:icross}. 
In all four ensembles, the slope of $e$ vs.~time
(top panel) is relatively shallow because
stirring and merging tend to counter-balance
each other. It is evident from the bottom panel that mergers begin 
immediately. Excepting the $i=0$ case, the inclination grows and 
reaches equipartition with eccentricity within a fraction of a Myr. 
As a result of this fast growth, the non-flat ensembles exhibit only
modest differences in their final $i$'s
and $\Delta$'s. As expected from Section 3.5, the flatter the system, the tighter the final spacing. The median final spacings are 20, 26, 27, and 29 from ensembles {\tt Eci0, Eci0.001, Eci0.1,} and {\tt Ec}, respectively.

\subsection{Larger initial eccentricity, wider final spacing}

Next we explore the effects of the initial eccentricity on the final spacing. Although $e$ does not enter directly into the spacing equilibrium derived in Section 3.5, it can have several indirect effects that we discuss below. For ensembles {\tt Ec, Ece0.3, Ece0.1}, and
{\tt E},
respectively,
$e_0 = 0.5 \, \delta a / a, 0.3 \, \delta a / a,
0.1 \, \delta a / a,$ and $0$;
these range from initially crossing orbits ({\tt Ec}) to initially circular orbits ({\tt E}).
For case {\tt E}, the initial 
inclination magnitude is drawn from a uniform distribution from 0--0.1$^\circ$,
while for the other cases $i_0 = e_0 / \sqrt{2}$ rad. All have the same
initial spacings $\Delta_0$.

\begin{figure*}
\begin{center}
\includegraphics{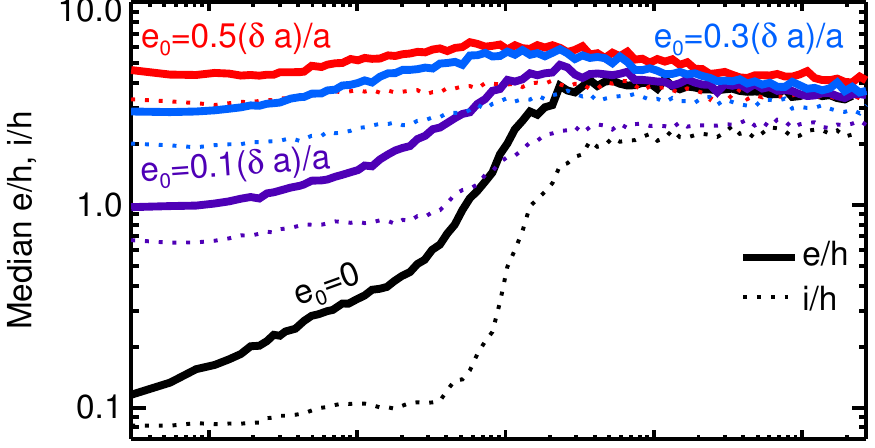} \includegraphics{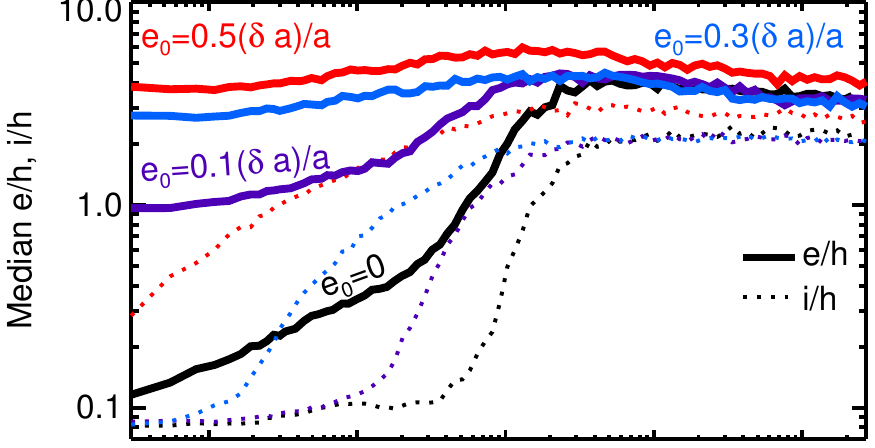}
\includegraphics{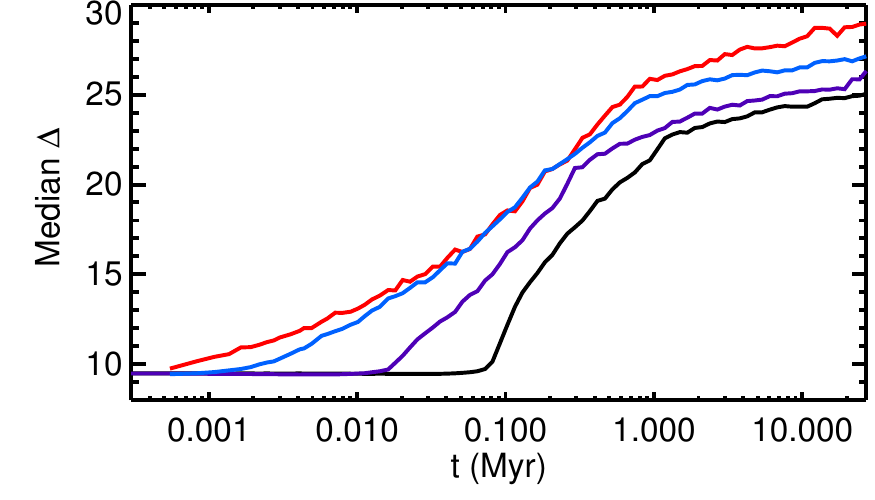}
\includegraphics{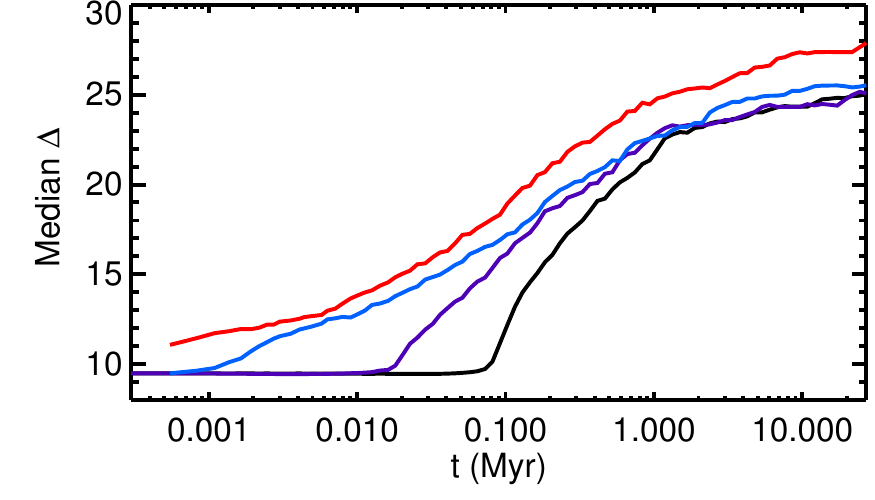}
\caption{
 \label{fig:cross} Initial eccentricities affect final spacings. Each of the ensembles (red, blue, purple, black) begins with identical spacings but different initial eccentricities.  Top: Evolution of $e$ (solid) and $i$ (dotted), averaged over all planets in an ensemble. The ensembles end up with similar final eccentricities but different inclinations. Bottom: Evolution of spacing $\Delta$. Left: Initial $i_0 = 
 e_0/\sqrt{2}$ rad. The larger the initial eccentricity, the wider the final spacing. Right: Initial $i_0$ magnitude drawn uniformly from 0--0.1$^\circ$. The final spacing differs only for the largest initial eccentricity (red).
}
\end{center}
\end{figure*}

We plot the evolution of $\Delta$, $e$, and $i$ in Fig.~\ref{fig:cross}, left panel. 
The black curve ($e_0=0$, {\tt E}) represents the same ensemble plotted as the red-dashed curve in Fig.~1. The purple curve ($e_0 = 0.1 \, \delta a /a$, {\tt Ece0.1}) begins with the
eccentricity growth stalled,
even though mergers are not yet occurring (bottom left panel).
It appears in individual simulations that the initial stalling occurs because we assigned random phases to the eccentricity vectors; orbits need to precess so
that their apsidal longitude differences $\Delta \varpi = \pi$ before encounters
become close and stirring begins in earnest.
The blue ($e_0 = 0.3 \,\delta a / a$, {\tt Ece0.3}) and red ($e_0 = 0.5 \, \delta a/ a$, {\tt Ec}) curves begin with nearly crossing or 
crossing orbits; as seen for the crossing orbits in Fig. 4, the counteracting contributions of mergers and stirring keep the eccentricity evolution slow.

Initial eccentricities affect final spacings: the median final spacings for the four ensembles ({\tt Ec, Ece0.3, Ece0.1, E}) are 29, 27, 26, and 25, respectively. Although the eccentricity does not explicitly factor into
the spacing equilibrium in Section 3.5, we hypothesize that it has the following indirect effects.

The first is that our initial conditions assumed equipartition between $e$ and $i$
($i_0 = e_0 / \sqrt{2}$).
The larger initial $e_0$ leads to a larger initial $i_0$ which propagates to a larger $i$ during orbit crossing, increasing the spacing required for an eccentricity equilibrium (Sections 3.5, 4.1). The ensembles that end up with wider spacings ({\tt Ec, Ece0.3}, red and blue, Fig.~5, bottom left panel) also have larger 
inclinations (top left panel). To test this idea further, we ran three additional ensembles of
simulations ({\tt Eci0.1, Ece0.3i0.1, Ece0.3i0.1}) that we plot along with ensemble {\tt E} in the right panels of Fig.~\ref{fig:cross}. For these auxiliary runs, instead of beginning with $i_0=e_0/\sqrt{2}$, we assign inclination magnitudes drawn from a uniform distribution from 0--0.1$^\circ$. 
Whereas the original ensembles {\tt Ec, Ece0.3, Ece0.1,} and {\tt E} (each with different $i_0$) converged to different final eccentricities, inclinations, and spacings (Fig.~5, left panel), the extra ensembles {\tt Eci0.1, Ece0.3i0.1, Ece0.1i0.1,} and {\tt E} converge to nearly identical final eccentricities and inclinations (right panel). The final spacing of ensemble {\tt Ece0.3i0.1} (26) differs only slightly from those of {\tt Ece0.1i0.1} and {\tt E} (25), despite their different initial eccentricities.

However, the ensemble with the largest initial eccentricities ({\tt Eci0.1}) ends up with larger final eccentricities, inclinations, and spacings. Therefore initial inclinations and equipartition are not the whole story. 
Two other effects may be contributing to large final spacings for the ensembles that began with large eccentricities  ({\tt Eci0.1, Ec}, $e_0 = 0.5 \, \delta a/a$, red curves). First, systems that start with eccentricities close to crossing may be prone to mergers that overshoot the spacing value corresponding to eccentricity equilibrium. Overshooting the benchmark spacing value is possible because collisions are non-reversible. Second, we assumed in Section 3 that relative velocities are shear-dominated. But when $e_0 = 0.5 \, \delta a / a$, epicyclic motions contribute significantly to relative velocities. Increasing the relative velocity reduces $\rgraze$, which reduces the merger rate and necessitates a wider spacing to achieve an eccentricity equilibrium.

\subsection{Smaller initial spacing, wider final spacing}

\begin{figure}
\begin{center}
\includegraphics{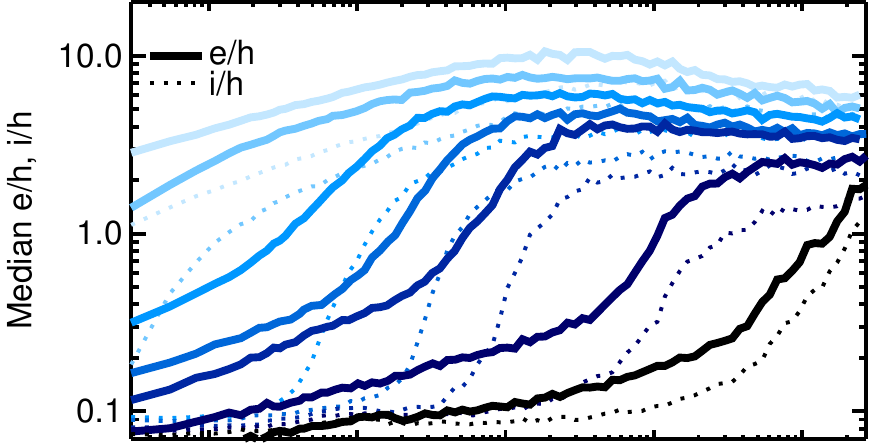}
\includegraphics{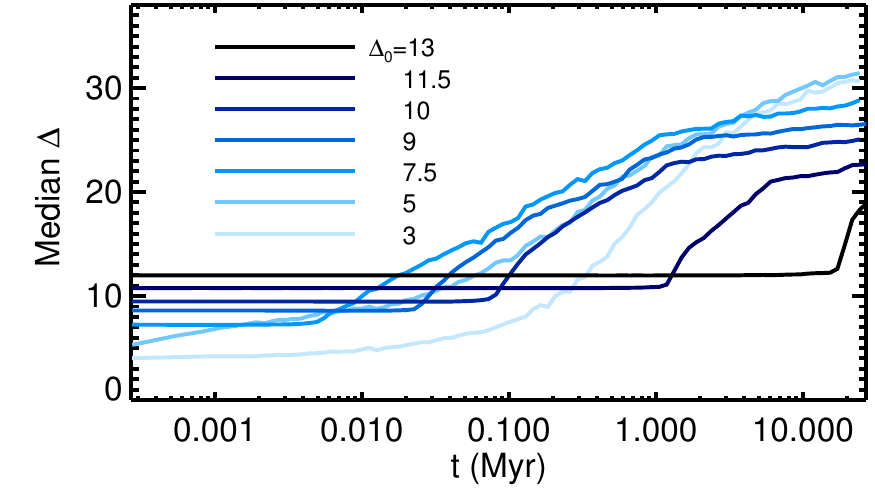}
\caption{ Initial spacings affect final spacings. Top: Evolution of $e$ (solid) and $i$ (dotted), averaged over all planets in an ensemble. Bottom: The tighter the initial spacing, the wider the final spacing.
 Initial spacings differ slightly from those listed
in the legend because of the way we lay down
our embryos; see the last paragraph of Section 2.
 \label{fig:rh}
}
\end{center}
\end{figure}

Last we explore the dependence of the final spacing on the initial
Hill spacing $\Delta_0$. We compare seven ensembles of
simulations: \{{\tt E$\Delta$3, E$\Delta$5, E$\Delta$7.5, E$\Delta$9,
  E, E$\Delta$11.5, E$\Delta$13}\}, which have initial $\Delta_0 =
\{3, 5, 7.5, 9, 10, 11.5, 13\}$, respectively. All have $e_0=0$ and
$i_0$ drawn from a uniform distribution from 0--$0.1^\circ$. We plot
the evolution of $e$, $i$, and $\Delta$ in Fig. \ref{fig:rh}. The
final median spacings of the ensembles are 31, 31, 29, 27, 25, 23, and
17, respectively. The systems that are initially spaced more tightly
end up with wider spacings.

Our interpretation is that the initially tighter spacings allow eccentricities to enter the orbit crossing regime sooner (i.e., at a tighter spacing),
further exciting eccentricities and inclinations (see Fig.~1). 
During the subsequent evolution, at a given $\Delta$, systems with initially smaller $\Delta_0$ have larger $i$ and $e$ than their counterparts that began with larger $\Delta_0$. As explored in Sections 4.1 and 4.2, larger $e$ and $i$ drive systems toward wider final spacings.

The equilibrium values of $e$ and $i$ achieved also seem to remember the initial spacing (Fig.~\ref{fig:rh}, row 1). The final equilibrium $e$ and $i$ stratify according to initial spacing: tighter initial spacings (light blue) result in higher final equilibrium $e$.

The assumed initial spacings of the $N$-body simulations presented here are intended to reflect possible endstates of oligarchic merging in the presence of residual gas and planetesimals. In the next section, we 
compute such endstates from scratch using
damped $N$-body simulations.

\section{Residual gas damping can establish  initial conditions that determine final spacings}
\label{sec:damp}

Before disk gas dissipates, it can damp
planet eccentricities and inclinations.
The competition between gas damping and
mutual viscous stirring by protoplanets
sets primordial orbital spacings,
eccentricities, and 
inclinations, which in turn determine 
the subsequent gas-free evolution.
We present here the results of $N$-body
simulations that include the effects of gas
damping.

\subsection{Damping prescriptions}
\label{sec:howdamp}

The strength and persistence of gas damping depends on how the protoplanetary nebula clears. Observations of young stellar clusters show that
transition disks --- disks with central 
clearings --- comprise $\sim$10\% of the 
total disk population \citep{Espa14,Alex14}.
We adopt the interpretation that all disks undergo
a transitional phase that lasts 
$\lesssim$ 10\% of the total $\sim$5--20 Myr
disk lifetime 
(\citealt{Koep13}; see also 
\citealt{Clar01}, \citealt{Drak09}, and \citealt{Owen11,Owen12} for 
their ``photoevaporation-starved'' model of disk evolution
which reproduces the short final clearing timescale).

For simplicity we model the gas disk as
having a fixed density that persists for 1
Myr and drops to zero thereafter. The step function could represent the
final Myr of the slow phase 
(prior to the formation of a central clearing)
followed by the fast transitional phase in which
gas is assumed to vanish completely inside 1 AU.
Alternatively the step function could represent
the act of clearing that takes place entirely during
the fast transitional phase.

We implement gas damping in a customized
version of {\tt mercury6} that incorporates
user-defined forces as
described in Appendix A of \citet{Wolf12},
adding several corrections and modifications
that allow us to use the hybrid symplectic
integrator, which we benchmarked against the
Burlisch-Stoer integrator. We impose $\dot{e}/e = -1/\tau$ and $\dot{i}/i = -2/\tau$ \citep{Komi02}.
Following \citet{Daws15}, we make use of 
three damping timescales $\tau$, based on the regimes identified in \citet{Papa00}, 
\citet{Komi02}, 
\citet{Ford07}, 
and \citet{Rein12}:
\begin{eqnarray}
\label{eqn:taudamp}
\tau = 0.003 \,d \left(\frac{a}{\rm AU}\right)^2 & \left( \frac{M_\odot}{M_{\rm p}} \right) \,{\rm yr} \,\times  \nonumber\\
& 1,& v < c_{\rm s} \nonumber \\
&\left(v/c_{\rm s}\right)^3,& v > c_{\rm s},\,\, i < c_{\rm s}/v_{\rm K} \nonumber  \\
&\left(v/c_{\rm s}\right)^4,& i > c_{\rm s}/v_{\rm K} \nonumber \\
\end{eqnarray}
\noindent where $M_\odot$ is one solar mass,
$M_{\rm p}$ is the planet mass, $v = \sqrt{e^2+i^2} v_{\rm K}$ is the random (epicyclic) velocity,
${v_{\rm K} = n a}$
with $n$ equal to the planet's mean motion,
${c_{\rm s} = 1.29 \,{\rm km/s} \,(a/{\rm AU})^{-1/4}} $ is the gas sound speed,
and $d$ is a constant proportional
to the degree of nebular gas depletion ($d=1$
corresponds approximately
to the minimum-mass solar nebula with
gas surface density $\Sigma_1 = 1700$ g/cm$^2$
at 1 AU, and $d>1$ corresponds to more depleted
nebulae).

Our simulations bear some
resemblance to those of
\citet{Komi02}, who also include
gas damping of planetary
random velocities. Our study
differs in having a 
damping timescale
that varies with epicyclic
velocity (Eqn.~\ref{eqn:taudamp});
solid surface densities $\Sigma_{z,1}$
that are up to an order of magnitude
larger than theirs (so as
to reproduce \kep super-Earths);
a gas disk evolution
that behaves as a step function
instead of their decaying exponential;
and hundreds more simulations
that enable us to make statements
with greater statistical confidence.
Our aim is also broader in that we
seek to understand
how spacings, inclinations,
and eccentricities inter-relate;
and later in Section 6 we will
bring planetary composition 
into this mix as we distinguish
between purely
rocky and gas-enveloped planets.
Where we overlap with \citet{Komi02},
we agree; in disks of low
$\Sigma_{z,1}$, producing
predominantly
rocky $<2 M_\oplus$ planets,
the final spacings $\Delta$
tend to be large, exceeding $\sim$20
(compare our Figure 8 with the
results described in the text of their
section 4).

\subsection{Conditions at the end of gas damping and subsequent evolution}

Ensembles {\tt Ed10$^0$, Ed10$^1$, Ed10$^2$, Ed10$^3$,} and {\tt Ed10$^4$} correspond to $d=1, 10, 10^2, 10^3,$ and $10^4$, respectively.
The simulations all have initial $\Delta_0 = 3$, $e_0=h$, $i_0 = 0.01 h$,
and $\rho=1$ g/cm$^3$.  The initial eccentricities are close to orbit crossing. The small initial inclinations are assumed to have been damped by the undepleted nebula prior to the start of the simulations.
Our choice of $i_0 = 0.01h$ gives inclinations
similar to those for $i_0 = 0.001^\circ$; the system is assumed to be
initially nearly flat.
Each ensemble comprises 80 simulations with solid surface densities
$\Sigma_{z,1}$ spanning 38--105 g/cm$^2$. Further details are listed in
Table \ref{tab:ens}. After 1 Myr, we shut off gas damping, and integrate for
an additional 27 Myr.

The physical interpretation of the scenario simulated here is as follows. Before the simulation begins, a high gas surface density has kept the embryos separated by $\Delta_0=3$. Then the gas density drops to the simulated depletion $d$, allowing the embryos to scatter and merge in the presence of gas damping for 1 Myr. Then the gas vanishes entirely and the system evolves for 27 Myr without gas. 

\begin{figure*}
\includegraphics{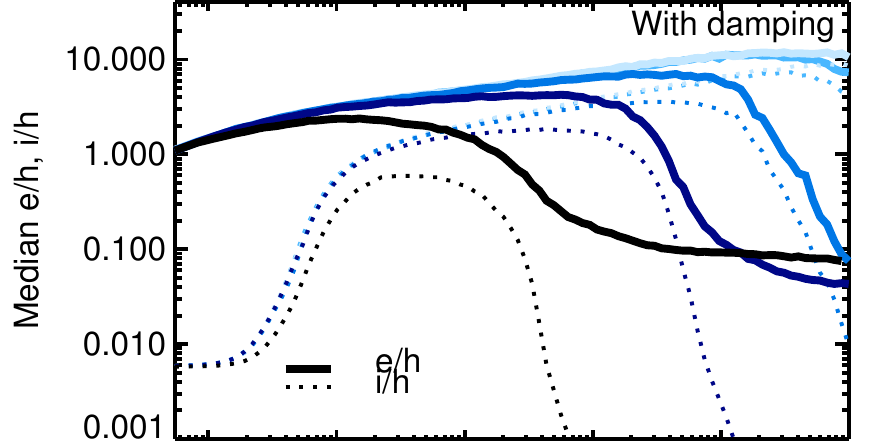}\includegraphics{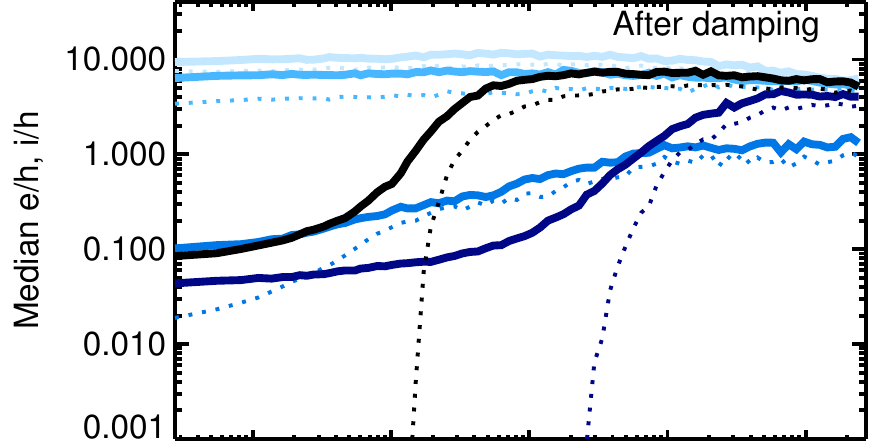}
\includegraphics{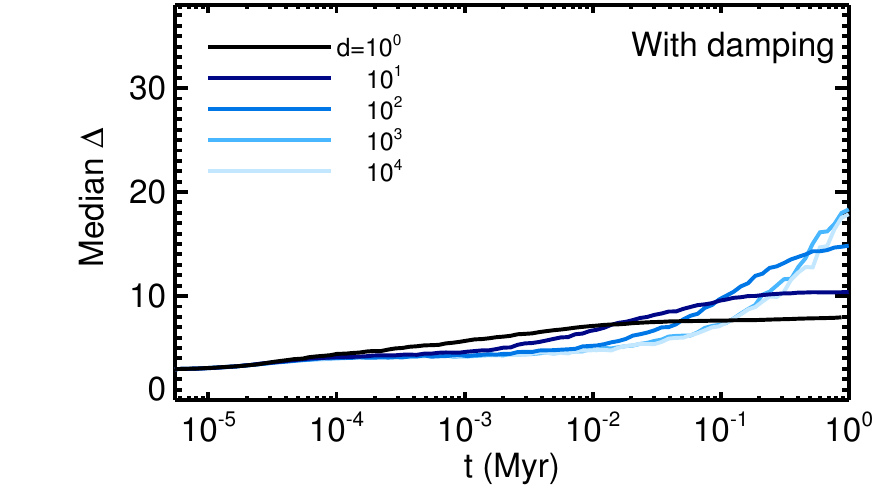}\includegraphics{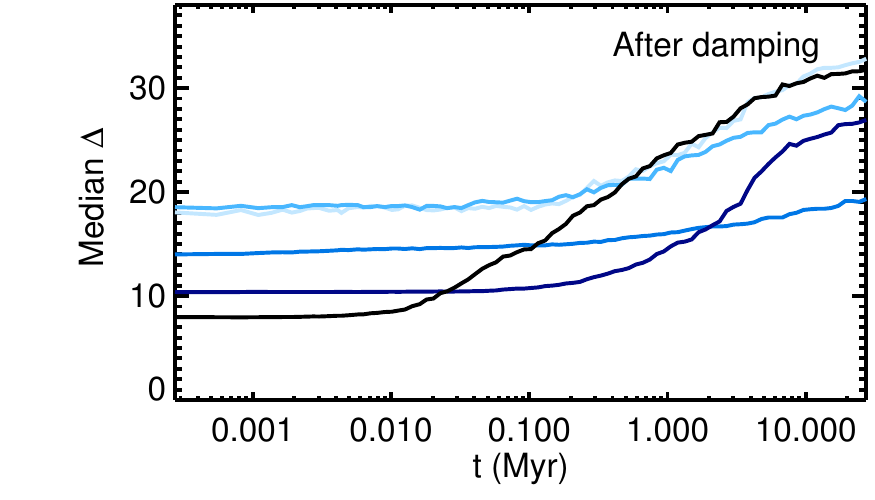}
\caption{ Evolution of $e/h$ and $i/h$ (top) and $\Delta$ (bottom) for ensembles of simulations with gas damping (left, parametrized by gas disk depletion
factor $d$; see text), and evolution after gas damping (right). 
 \label{fig:damp}
}
\end{figure*}

The left panels of Fig.~\ref{fig:damp} show the evolution of $i$, $e$, and $\Delta$ in the presence of damping. Gas damping flattens and circularizes systems to low $i$ and $e$ on  timescales that increase with $d$ (top left).  After 1 Myr of evolution with gas damping, the smallest $d$ (strongest damping, dark blue) ensembles have the tightest spacings, smallest inclinations, and smallest eccentricities
(the $d=1$ ensemble is a bit of an outlier
in eccentricity). The subsequent undamped,  gas-free evolution is shown in the right panels.

In the more strongly damped {\tt Ed$10^0$}, {\tt Ed$10^1$}, and {\tt Ed$10^2$} ensembles, the eccentricity and inclination (Fig. \ref{fig:damp}, top left) initially grow. The initial eccentricities are close to orbit crossing ($e_0/h = 1 = \Delta_0$/3) and so  embryos begin to merge almost immediately (bottom left). As the spacing increases, eccentricity and inclination growth rates decrease and eventually $e$ and $i$ are damped. Inclinations are damped sooner and more steeply. Eccentricities are damped later and appear to plateau to fixed points. When damping ends, the ensembles with the shortest damping timescales have the tightest spacings, but eventually their fortunes will reverse. As documented in Section 4.3, the tightest initial spacings yield the widest final spacings in the gas-free stage. Among the {\tt Ed$10^0$}, {\tt Ed$10^1$}, and {\tt Ed$10^2$} ensembles, all of which emerge from the damping stage with similarly small eccentricities, the {\tt Ed$10^0$} ensemble ends up most widely spaced in the subsequent gas-free evolution, while {\tt Ed$10^2$} ends up most tightly spaced. We highlight the {\tt Ed$10^2$}  ensemble as producing typical final spacings of $\Delta \sim 20$ (Fig.~7, bottom right panel) in agreement with the observations (e.g., \citealt{Fang13}). We will compare to the observations in greater detail in Section 7.

The less-damped ensembles {\tt Ed$10^3$} and {\tt Ed$10^4$} reach the widest spacings. At the end of gas damping, they begin (right panels) on initially nearly crossing orbits, like those explored in Section 4.2. In their subsequent evolution, they reach a spacing of $\Delta \approx 30$. 

We also see the effects of the 
inclinations
during orbit crossing, as described in Section 4.1. During the gas-free evolution, the {\tt Ed$10^1$} ensemble ends up with an average final eccentricity similar to those of the {\tt Ed$10^3$} and {\tt Ed$10^4$} ensembles, but the
inclinations
of {\tt Ed$10^1$} when orbits begin to cross ($t \sim 1$ Myr in the right panel) are low. The lower 
inclinations
likely contribute to the narrower final spacing of {\tt Ed$10^1$} compared to {\tt Ed$10^3$} and {\tt Ed$10^4$}.

We continued integrations of three ensembles ({\tt Ed$10^1$, Ed$10^2$, Ed$10^3$}) up to a total time of 300 Myr (corresponding to four months of wall-clock time) and found that their outcomes hardly changed from those shown in Fig.~7. The eccentricities $e/h$ of these three
ensembles remained approximately constant, while their  median $\Delta$ values grew by about 10\%.  We estimate
that the median $\Delta$ may grow by another $\sim$10--15\% if the integrations were extended to 10 Gyr. The {\tt Ed$10^2$} ensemble retained its much tighter spacings and smaller eccentricities and inclinations as compared to the other two ensembles.  Although the ensembles retained their qualitative character when integrated
for longer, the quantitative changes,
in particular to the spacings,
can be significant when attempting
to fit to the observations (cf.~\citealt{Pu15}).
Longer integrations of all ensembles could be run in a follow-up study. 

\section{Spacing and planetary bulk density}
\label{sec:comp}

Here we demonstrate several ways in which the giant impact stage explored in Sections 3--6 connects planets' bulk densities to their orbital properties. More rarefied planets are characterized by tighter orbital spacings, lower eccentricities, and lower mutual inclinations. Our results motivate a search for observational correlations between super-Earths' orbital and compositional properties, which statistical studies usually assume to be independent of each other (e.g., \citealt{Fang12b}).

\subsection{High solid surface density $\rightarrow$ Rarefied and tightly spaced planets}

\begin{figure}
\begin{center}
\includegraphics[width=\columnwidth]{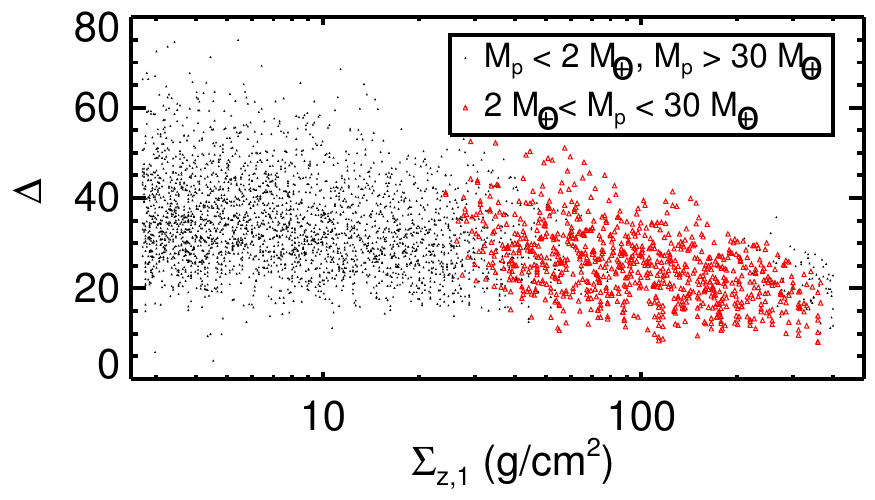}
\caption{Dependence of final spacings $\Delta$ on surface density normalization $\Sigma_{z,1}$ from ensemble {\tt Eh}. Red points are planets with $2 M_\oplus < M_p < 30 M_\oplus$, and black points fall 
outside this mass range. The spacing $\Delta$ tends to decrease with $\Sigma_{z,1}$, particularly for a restricted range of masses (e.g., the red points). The trend is not sensitive to the initial conditions of the ensemble. 
 Our result that
$\Delta \gtrsim 20$
for low $\Sigma_{z,1}$ agrees with
\citet{Komi02}.
\label{fig:sig}
}
\end{center}
\end{figure}

Planets with lower bulk densities and planets with tighter spacings are each a consequence of higher solid surface density disks. In a higher solid surface density disk, a core of a given mass can form from mergers of embryos contained in a narrower zone of smaller $\Delta$. Figure \ref{fig:sig} demonstrates the anti-correlation between the final spacing of super-Earths and the solid surface density normalization. Furthermore, when cores grow from narrower feeding zones, they undergo mergers quickly (see also the empirical study of orbit crossing timescales as a function of $\Delta$ by \citealt{Yosh99}), reaching a mass large enough to acquire gas envelopes before the gas disk dissipates.

\begin{figure}
\begin{center}
\includegraphics{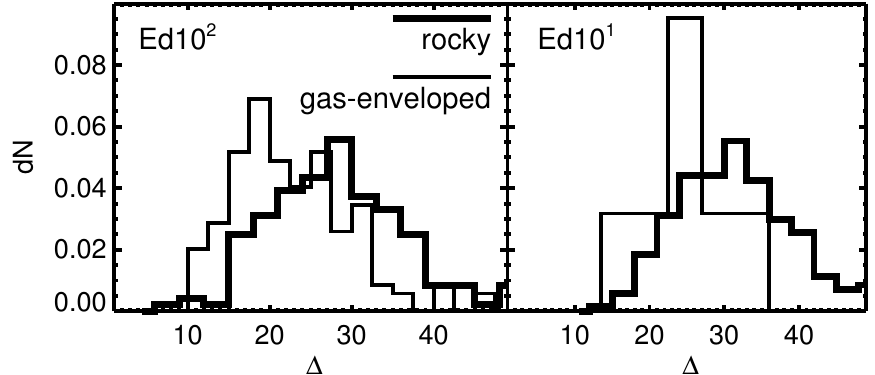}
\caption{Histograms of spacings $\Delta$ for simulated rocky planets (thin) and gas-enveloped (thick) from {\tt Ed$10^2$} (left) and {\tt Ed$10^1$} (right). Planets included have masses ranging from 2--30 $M_\oplus$. Planet pairs are labeled gas-enveloped when both members of the pair have accreted 1\%-by-mass atmospheres within 1 Myr; all other pairs are labeled rocky. The gas-enveloped planet pairs have tighter spacings. \label{fig:deltadensesim}
}
\end{center}
\end{figure}

Following \citet{Daws15}, we apply the \citet{Lee14} nebular accretion models to track the gas fractions of planets as they
grow through mergers in the {\tt Ed$10^1$} and {\tt Ed$10^2$} ensembles, labeling planets that acquire a 1\% atmosphere in less than 1 Myr as ``gas-enveloped,''  and planets that do not as ``rocky.'' We compute the final spacing between each pair of adjacent planets and plot the resulting spacing distributions in Fig.~\ref{fig:deltadensesim}. Gas-enveloped planets end up more tightly spaced than rocky planets. 
To test whether these differences are statistically
significant, we apply a two-sample Kolmogorov-Smirnov (K-S) test to the gas-enveloped vs.~rocky distributions of $\Delta$ from {\tt Ed$10^2$}  (Fig.~\ref{fig:deltadensesim}, left panel), obtaining a $p$-value of $5\times10^{-9}$. For the distributions from {\tt Ed$10^1$} (right panel), we obtain
$p = 1.1\times10^{-7}$. These $p$-values are sufficiently
small that for both the
{\tt Ed$10^2$} and {\tt Ed$10^1$} ensembles, we reject
the null hypothesis that the gas-enveloped and rocky planet
spacings are drawn from the same underlying distribution.

In the {\tt Ed$10^2$} ensemble, about half of pairs with masses $2 M_\oplus < M_p < 30 M_\oplus$ consist of two gas-enveloped planets. These pairs have lower median eccentricities ($\tilde{e} = 0.02$) and
inclinations ($\tilde{i} = 0.8^\circ$) than pairs with one or more rocky planets ($\tilde{e} = 0.06, \tilde{i} = 2.5^\circ$). 
 We apply a two-sample K-S test to the gas-enveloped vs.~rocky distribution of $e$ ($i$) and obtain a $p$-value of $1.1\times10^{-7} (6\times10^{-6})$. We therefore reject the null hypothesis that the gas-enveloped and rocky planet
eccentricities (inclinations) are drawn from the same $e$ ($i$) distribution.

\subsection{Moderate gas damping $\rightarrow$ Lower bulk densities and tighter spacings}

We showed in Section 5 that a depletion factor $d \sim 10^2$,
corresponding to $\Sigma_{{\rm gas},1} = 17$ g cm$^{-2}$, produced the
tightest final spacings.\footnote{A
  depletion factor of $\sim$$10^2$ relative to the MMSN corresponds to
  a factor of $\sim$$10^3$ depletion relative to the minimum-mass
  extrasolar nebula constructed by \citet{Chia13}.} The planets in the {\tt Ed10$^2$} ensemble
undergo most of their growth during the gas damping stage, as can be
seen in Fig.~\ref{fig:damp}: $\Delta$ grows most during the gas
damping stage (left panel) rather than the gas-free stage (right
panel). For the range of solid surface densities explored in this
ensemble (38--105 g cm$^{-2}$, Table 1), the gas-to-solid ratio is
$\sim$0.2--0.4, so there is sufficient disk gas
locally for planets to acquire low-mass
(1\%) atmospheres without having to appeal
to gas accreting inward from outside 1 AU
\citep[cf.][]{Lee16}.

\begin{figure}
\begin{center}
\includegraphics{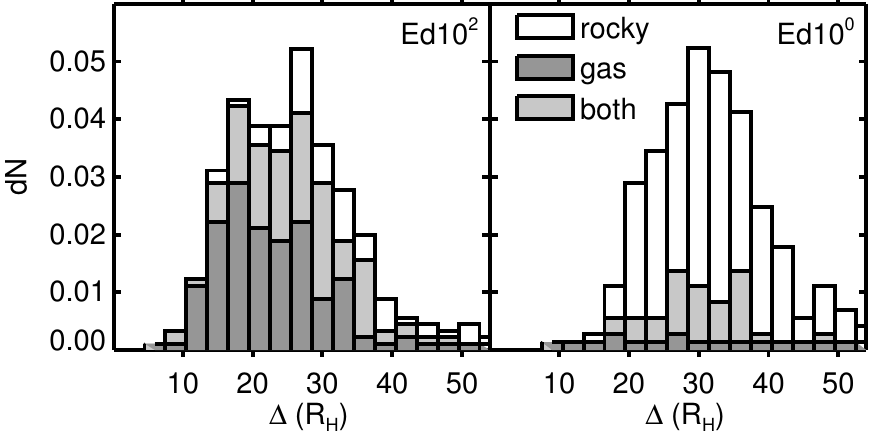}
\caption{Stacked bar charts (not histograms lying behind one another) of spacings $\Delta$ for simulated rocky pairs (no shading), gas-enveloped pairs (dark shading), and pairs of one rocky and one gas-enveloped planet (light shading) from the {\tt Ed$10^2$} (left) and {\tt Ed$10^0$} (right) ensembles. Planets included have masses ranging from 2--30 $M_\oplus$. The {\tt Ed$10^2$} ensemble has a much greater fraction of gas-enveloped planet pairs than the {\tt Ed$10^0$} ensemble.
\label{fig:bothcomp}
}
\end{center}
\end{figure}

The simulations
with depletion factors that yield wider spacings ($d=1, 10, 10^3, 10^4$)
also yield higher density planets (as illustrated for $d=1$ compared to $d=10^2$ in Fig. \ref{fig:bothcomp}). On the one hand,
when the depletion factor is small ($d=1, 10$), the planets are
largely assembled after the gas disk completely dissipates
(Fig.~\ref{fig:damp}, bottom right panel): they end up with
lower gas fractions and higher densities.
On the other hand, when the
depletion factor is much higher (gas-to-solid ratios of
$\sim$0.02--0.04 when $d=10^3$ or $\sim$0.002--0.004 when $d=10^4$),
then there is insufficient gas during the growth stage to create significant
atmospheres
(assuming gas in the inner disk where cores reside
is not replenished from the outer disk; 
see \citealt{Lee16} who relax this assumption).
The case $d=10^2$ is a happy middle ground for making
low-density planets.

\begin{figure*}
\begin{center}
\includegraphics{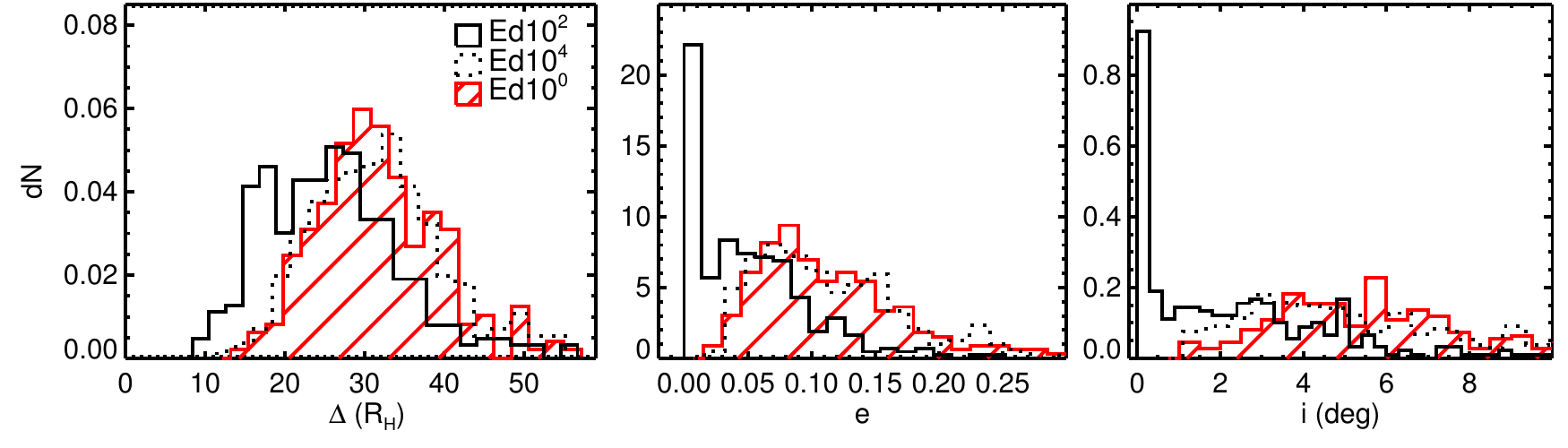}
\caption{ 
Final distributions (at the end of 28 Myr) of $\Delta$, $e$, and $i$ for planets with $2 M_\oplus < M_{\rm p} < 30 M_\oplus$ from three ensembles: {\tt Ed10$^0$,Ed10$^2$,Ed10$^4$}. Ensemble {\tt Ed10$^2$}, which has moderate gas damping during the first 1 Myr, exhibits the tightest spacings and smallest eccentricities and inclinations.
\label{fig:deltadamp}
}
\end{center}
\end{figure*}

Median eccentricities and inclinations in the most tightly spaced {\tt
  Ed10$^2$} ensemble tend to be low: $\tilde{e} = 0.04$ and $\tilde{i} =1.8^\circ$
for $2 M_\oplus < M_{\rm p} < 30 M_\oplus$. In contrast,
$\tilde{e} = \{ 0.10, 0.08,0.08,0.10\}$ and $\tilde{i}= \{5, 4, 4, 6^\circ \}$
for ensembles \{{\tt Ed$10^0$, Ed$10^1$, Ed$10^3$, Ed$10^4$}\},
respectively. Figure \ref{fig:deltadamp} compares the
final spacing, eccentricity, and inclination distributions for three
ensembles ({\tt Ed10$^0$}, {\tt Ed10$^2$}, and {\tt Ed10$^4$}).  
 We apply two-sample K-S tests to test whether
the differences between these three ensembles are
statistically significant. We reject the null hypotheses that 
the $\Delta$, $e$, and $i$ values for ensemble {\tt Ed10$^2$}
are drawn from the same distributions as for {\tt Ed10$^4$}
or {\tt Ed10$^0$}; the $p$-values are less than $2\times 10^{-7}$.
In contrast, we cannot reject the null hypotheses that {\tt Ed10$^0$} and {\tt Ed10$^4$} are drawn from the same distribution with
respect to $\Delta$ ($p =0.95$) and $e$ ($p=0.91$). For $i$, we reject the null hypothesis that {\tt Ed10$^0$} and {\tt Ed10$^4$} are drawn from the same distribution ($p =0.003$).

\subsection{Larger collisional cross section (lower bulk density) $\rightarrow$ Tighter spacing}

\begin{figure}
\begin{center}
\includegraphics[width=\columnwidth]{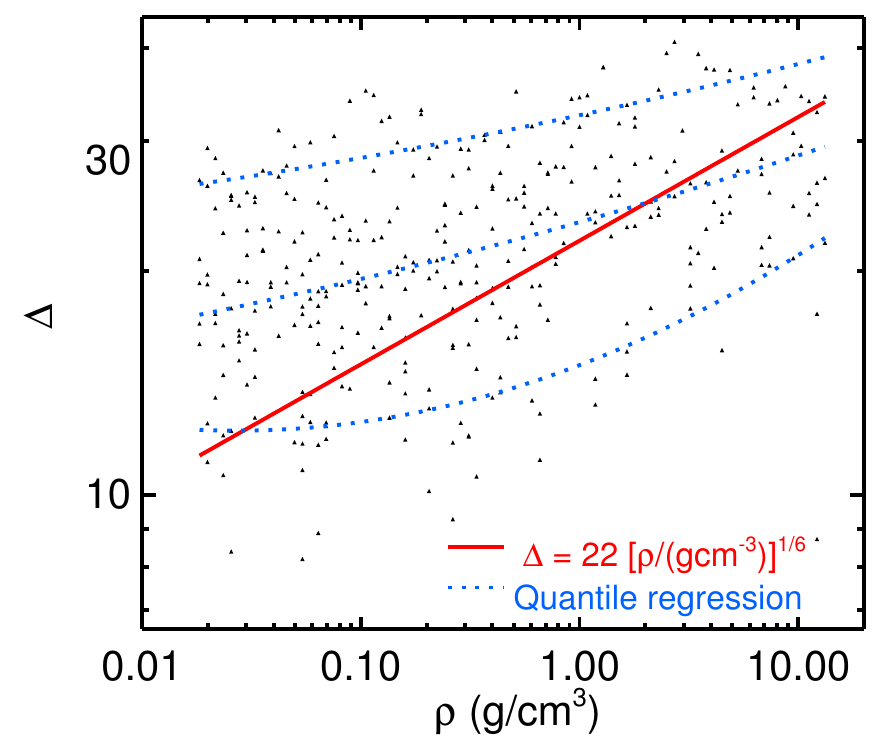}
\caption{Spacing $\Delta$ vs.~planet bulk density
$\rho$ from the {\tt Eh$\rho$} suite of simulations, where $\rho$ is the assumed constant bulk density that we assign
and which defines the collisional cross section.
The red line is the $\Delta = 22 \left({\rho}/{\rm g\,cm^{-3}}\right)^{1/6}$ scaling expected from Eqn.~\ref{eqn:delta}.
The blue lines model the upper envelope (90\%), median, and lower envelope (10\%) of the simulation data, computed using quantile regression. The red line is steeper than the upper envelope and median, but approximates the slope of the lower envelope for $\rho>1$ g/cm$^3$.
\label{fig:dense} }
\end{center}
\end{figure}

From the order-of-magnitude arguments in Section 3, we expect the spacing to scale approximately with
planet bulk density as $\rho^{1/6}$ (Equation \ref{eqn:delta}). The
dependence arises because the planet's size determines its collisional
cross-section, which in turn affects the balance between mergers and scatterings and
therefore the spacing required for an eccentricity
equilibrium. Although the scaling of spacing with bulk density is
weak, \kep super-Earths span more than a factor of 10 in bulk
density (e.g., \citealt{Cart12,Wu13,Hadd14,Weis14}). We perform a
suite of 80 simulations ({\tt Eh$\rho$}) spanning a range of $\rho$
from 0.02 g/cm$^3$ to 14 g/cm$^3$ (these are more extreme than those observed)
and plot the dependence of $\Delta$ on $\rho$ in
Fig.~\ref{fig:dense}. 
 Overplotted in red is a 
$\Delta \propto \rho^{1/6}$ line for comparison with the upper envelope (90\%), median, and lower envelope (10\%)  of the simulation points (blue lines; these are computed by quantile regression using the {\tt COBS} package in {\tt R}; see \citealt{Ng07,Ng15,Rcor15}). The red line is steeper than the upper envelope and median, but nearly matches the slope of the lower envelope for $\rho>1$ g/cm$^3$.
The upper envelope may reflect systems that overshoot their equilibrium spacings; since all {\tt Eh$\rho$} runs start with the same initial
eccentricities and spacings, lower $\rho$ systems start closer to their
equilibrium spacings and are therefore more prone to overshooting.
We conclude that smaller $\rho$ indeed produces tighter spacings.
Planets with larger $\rho$ also have larger median 
$e$ and $i$: $\tilde{e} = 0.06$ and $\tilde{i}=1.9^\circ$ for planets with
$\rho < 1$ g cm$^{-3}$ vs. $\tilde{e} = 0.10$ and $\tilde{i}=5.2^\circ$ for planets
with $\rho >1$ g cm$^{-3}$.

The dependence of spacing on collisional cross section motivates a more realistic treatment of the planets' compositional and collisional evolution. Here we assumed that the bulk density remains constant as the planets grow from embryos to super-Earths, but ideally simulations should account for the change in density as material is compacted and atmospheres are accreted or eroded. The collisional prescription used here is overly simplistic and does not account for the internal structure of the planet: some or all of the low-density planets from \kep are thought to be rocky cores with rarefied atmospheres and hydrodynamic simulations are needed to accurately assess the collisional outcomes, including the dependence on the collisional impact parameter.

\subsection{Shorter orbital periods $\rightarrow$ Wider spacings and higher bulk densities}

\begin{figure}
\begin{center}
\includegraphics[width=\columnwidth]{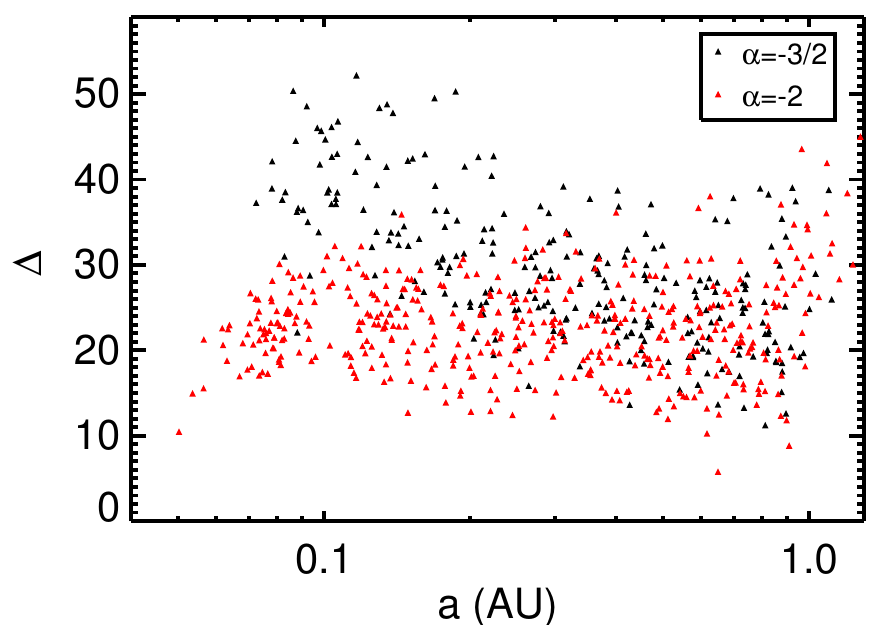}
\includegraphics[width=\columnwidth]{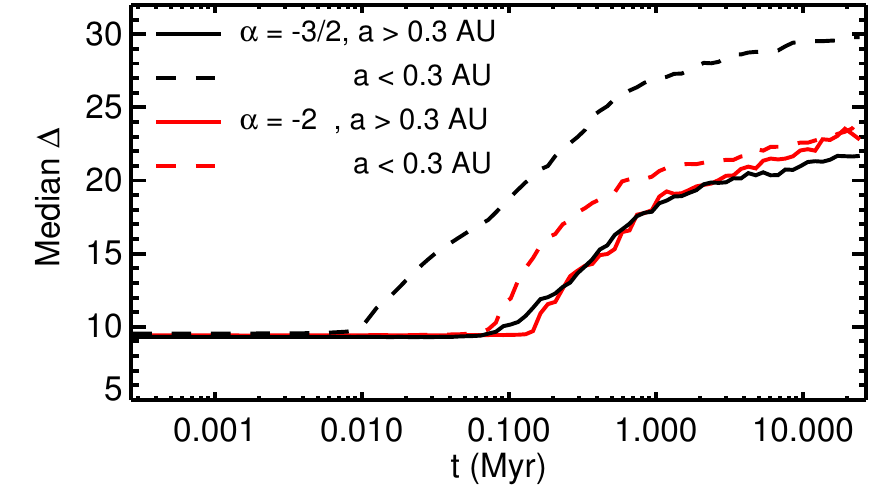}
\caption{Top: Spacing $\Delta$ vs.~semi-major axis from simulations with our default surface density profile $\Sigma_z = \Sigma_{z,1} (a/{\rm AU})^{-3/2}$ (black, ensemble {\tt Eh}) and with a steeper profile $\Sigma_z = \Sigma_{z,1} (a/{\rm AU})^{-2}$ (red, ensemble {\tt Eh$\alpha$-2}) that results in a constant isolation mass with semi-major axis (Eqn.~\ref{eqn:memb}). For $\alpha=-3/2$, $\Delta$ decreases with semi-major axis; for $\alpha=-2$, $\Delta$ is constant with semi-major axis. Bottom: Time evolution of average $\Delta$ for $\alpha=-3/2$ (black) and $\alpha=-2$ (red) for $a < 0.3$ AU (dashed) and $a > 0.3$ AU (solid). The evolution differs between the $\alpha=-3/2$ and $\alpha=-2$ ensembles in the inner disk ($a < 0.3$ AU) but not in
the outer disk ($a>0.3$ AU).
\label{fig:per}
}
\end{center}
\end{figure}

From Eqn.~\ref{eqn:delta}, we expect tighter spacings at smaller orbital periods. However, when we plot the final $\Delta$ vs. semi-major axis in Fig. \ref{fig:per} (using ensemble {\tt Eh}), we see the opposite trend: wider spacings at smaller semi-major axes. We hypothesize that this behavior arises
because of the particular solid surface density power law we have chosen. For $\alpha=-3/2$ (Section 2),
embryo masses increase with $a$, resulting in
more massive distant planets. These distant
planets gravitationally stir closer-in planets,
necessitating wider spacings to achieve an
eccentricity equilibrium.

To test this hypothesis, we run an additional ensemble of simulations ({\tt Eh$\alpha$-2}; Table 1) with  $\alpha=-2$, resulting in an embryo mass that is
constant with semi-major axis. We find spacings similar to those of the {\tt Eh} ensemble at longer orbital periods, but tighter spacings than the {\tt Eh} ensemble at shorter orbital periods.
In the bottom panel of Fig.~\ref{fig:per}, we plot the time evolution of the median $\Delta$. For $a > 0.3$ AU, the evolution proceeds similarly regardless of $\alpha$. For $a<0.3$ AU, the mergers begin earlier in the $\alpha=-3/2$ ensemble, implying that eccentricities are more quickly excited and orbits
cross earlier. All of these results are consistent with our hypothesis that massive planets in the outer disk can contribute significant ``non-local'' stirring, particularly for
shallow surface density profiles.

The results of this subsection are consistent with a recent study by \citet{Mori15} who simulated the growth of super-Earths in disks with $\{\alpha = -5/2, -3/2, -1/2\}$. They found tighter spacings, higher multiplicities, and smaller eccentricities and inclinations for planets formed in steeper $\alpha$ disks. The differences are most
marked for the shortest period planets (see their Figs.~3 and 4).

In summary, for disks with solid surface density profiles less steep than
$\alpha=-2$, planets at shorter orbital periods form denser and more widely
spaced. For the $\alpha=-3/2$ profile used throughout this work,
shorter period 
planets tend to have smaller core masses; they are expected to accrete
less gas from the nebula and to more easily lose what little envelopes
they may have gathered to photoevaporation. Short period planets also tend to have larger median eccentricities and inclinations ($\tilde{e}=0.08$ and $\tilde{i}=4^\circ$ for $a<0.3$ AU in the {\tt Eh} ensemble) than do longer period planets ($\tilde{e}=0.05$ and $\tilde{i}=2^\circ$ for $a>0.3$ AU).

\section{Comparison to \kep multi-tranet systems}

\begin{figure*}
\begin{center}
\includegraphics[width=\columnwidth]{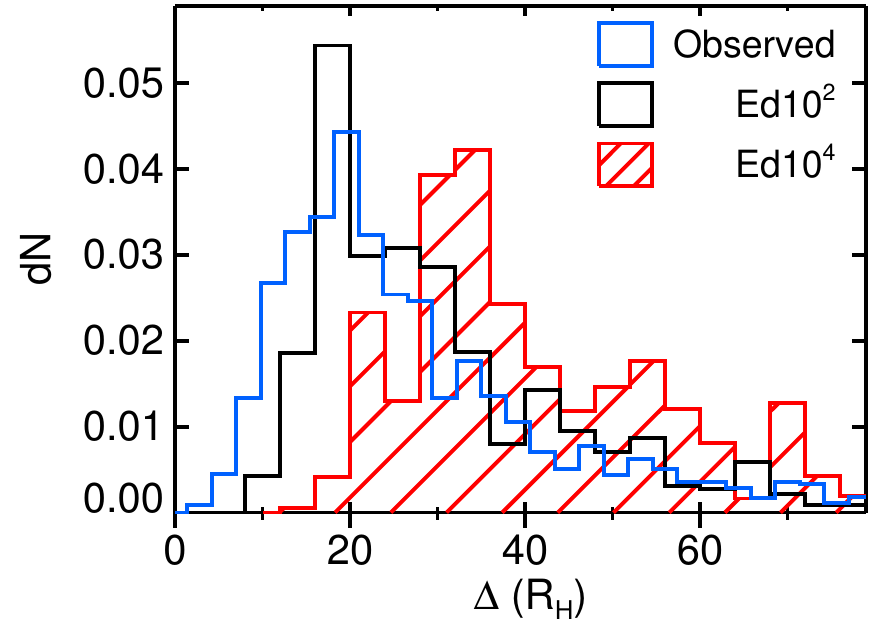}
\includegraphics[width=\columnwidth]{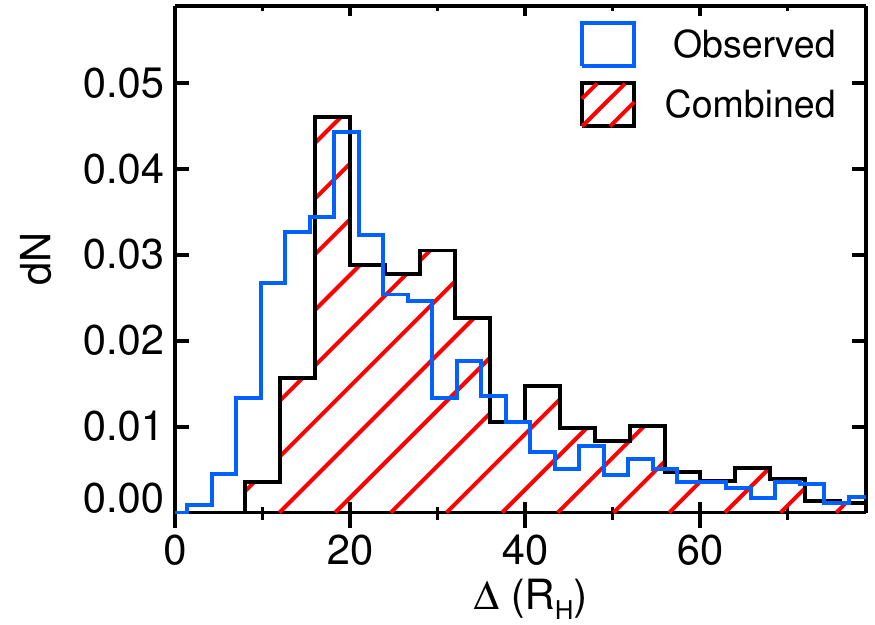}
 \includegraphics[width=\columnwidth]{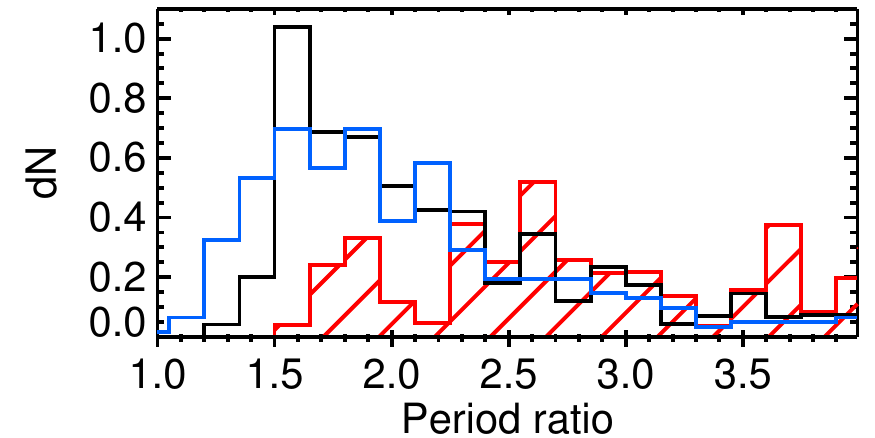}
  \includegraphics[width=\columnwidth]{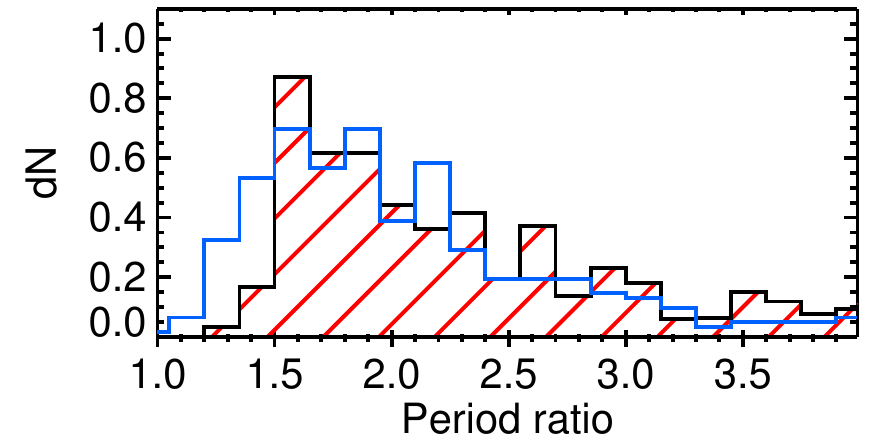}
 \includegraphics[width=\columnwidth]{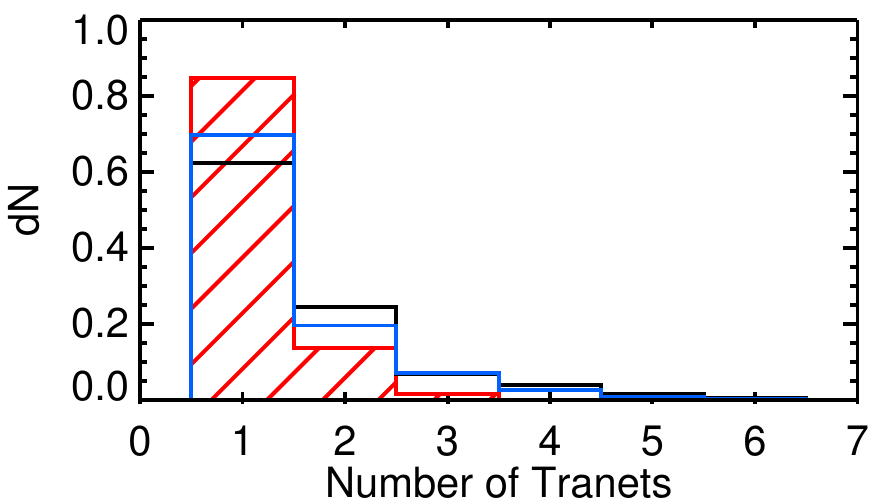}
 \includegraphics[width=\columnwidth]{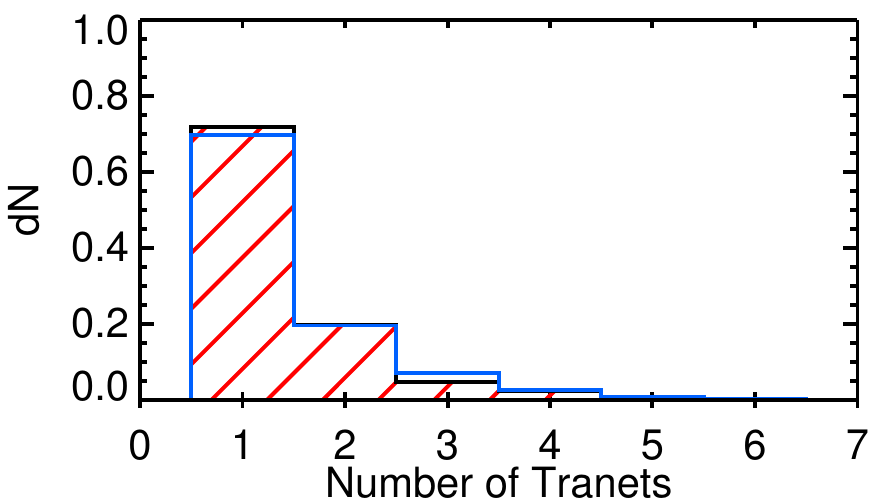}
 \includegraphics[width=\columnwidth]{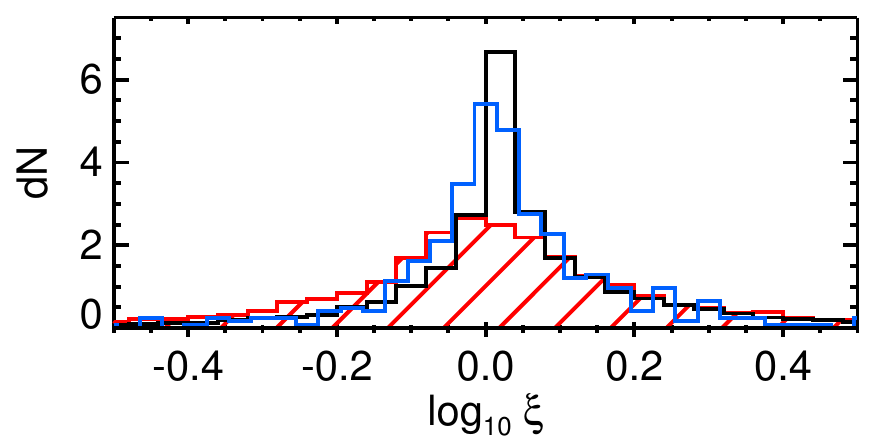}
\includegraphics[width=\columnwidth]{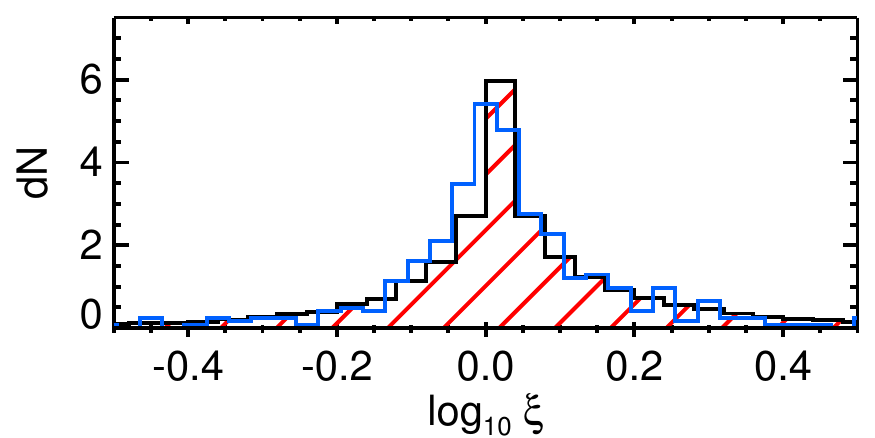}
\caption{
\label{fig:tranet} Properties of tranets generated from simulations. Row 1: Distribution of Hill spacing $\Delta$, Row 2: Period ratio, Row 3: Number of tranets per system, Row 4: Transit duration ratio. Left: Distributions from ensembles {\tt Ed10$^4$} (red) and {\tt Ed10$^2$} (black) compared to observations (blue). Right: Distributions from a mixture of the two 
ensembles
(roughly 1:2 for tranets from {\tt Ed10$^4$}:{\tt Ed10$^2$}).
The mixture matches the observed distributions well except that it is missing the tightest spacings and smallest period ratios.
}
\end{center}
\end{figure*}

We make some basic comparisons to the \kep sample to assess how the conditions for late-stage planet formation explored in Sections 3--6 manifest in \kep observables. We convert our simulated planets to tranets by imposing a rough sensitivity cut of $M_p > 2 M_\oplus$ and $P<200$ days. For each simulated system, we generate $10^4$ systems randomly oriented in space
with respect to an observer.
The subset of planets that transit comprise our tranet sample. We plot spacings from the {\tt Ed10$^2$} and {\tt Ed10$^4$} ensembles in
Fig.~\ref{fig:tranet}, row 1, left panel. Compared to Fig.~\ref{fig:deltadamp} (intrinsic
planet spacings), observational
selection effects slightly decrease the mode of the distribution of tranet spacings and enhance the tail.

To compare our synthetic $\Delta$ distribution
with that based on observations, we queried the cumulative \kep candidates from the NASA Exoplanet Archive (March 16, 2015), restricting our sample to host stars with ${4100 {\rm K}  < T_{\rm eff} <  6100 {\rm K}}$, $4 < \log g < 4.9$, and \kep magnitude $< 15$, and 
selecting for systems in which at least one planet has $R < 4 R_\oplus$. Statistical modeling of the potential astrophysical false positive population has revealed that less than 10\% of the candidates are false positives \citep{Mort11,Fres13}, allowing us to treat the sample as representative of true planets. We use the \citet{Wolf15} probabilistic mass-radius relation to convert the observed candidate periods to a $\Delta$ distribution. The resulting distribution is not sensitive to the exact mass-radius relation used \citep{Fang13}.\footnote{We use the same mass-radius relationship for planets with different period ratios and spacings despite our expectation from Section 6 that more tightly spaced planets have lower densities.}  The observed spacing distribution has a
mode $\Delta \sim 18$,
similar to our most tightly spaced distributions (e.g., {\tt Ed10$^2$} from our ensembles that include a gas damping stage), but also
has a tail more consistent with our wider-spaced distributions.

Following \citet{Hans13}, we compare several other properties of our simulated tranets to the observed \kep planets. In the left panels of Fig.~\ref{fig:tranet}, we compare period ratio (row 2), number of tranets per system (row 3), and transit duration ratio (row 4) distributions. The transit duration ratio of a pair of planets (the inner
labeled ``1'' and the outer labeled ``2'') is:
\begin{equation}
\xi = \left(\frac{1-B_1^2}{1-B_2^2}\right)^{\frac{1}{2}}\left(\frac{1-e_2^2}{1-e_1^2}\right)^{\frac{1}{2}}\left(\frac{1+e_1\sin\omega_1}{1+e_2\sin\omega_2}\right)
\end{equation}
\noindent where $B$ is the transit impact parameter and $\omega$ the argument of periapse. As argued by \citet{Fabr14}, the duration ratio distribution gives insight into the mutual inclinations. In flatter systems, the inner planet in a pair tends to have a smaller impact parameter, skewing the $\xi$ distribution to values greater than 1. Both the mutual inclinations and eccentricities affect the width of the $\xi$ distribution, but mutual inclination has a stronger effect because a small change in inclination has a strong effect on $B$,
whereas a small change in $e$ affects $\xi$
only modestly.

The ensembles that produce the widest spacings (represented by {\tt Ed10$^4$} in Fig.~\ref{fig:tranet})
also produce the largest period ratios, the lowest multiplicities, and the widest/least-skewed $\xi$. They overproduce single tranets and underproduce high multiplicity systems. Their spacing, period ratio, and $\xi$ distributions match the tails of the observed distributions but not the peak. The \citet{Hans13} initial conditions also result in spacings, period ratios, and $\xi$ distributions that are too broad.\footnote{Although \citet{Hans13} report an underproduction of single tranets, we find an excess of single tranets for our ensemble of simulations using their initial conditions (not shown).}

In contrast, the ensembles that produce the tightest spacings (exemplified by the intermediate damping ensemble, {\tt Ed10$^2$},
in Fig.~\ref{fig:tranet}) match the peaks of the observed spacing and period ratio distributions but not the tails. They overproduce high multiplicity systems and underproduce single tranets. They produce a $\xi$ distribution that is too narrow and skewed.

Rather than resembling a single one of our simulated ensembles, the observed distributions appear to comprise a mixture of ensembles. The simulated  distributions from one ensemble can either match the peaks or tails of the observed distributions, but not both. In the right column of Fig.~\ref{fig:tranet}, we combine the tranets from a tightly spaced, low mutual inclination ensemble ({\tt Ed$10^2$}) with a widely spaced, larger mutual inclination ensemble ({\tt Ed$10^4$}). For the combined distribution, we use all the {\tt Ed$10^2$} tranets and randomly draw 50\% of the generated {\tt Ed$10^4$} tranets (where the 50\% weighting is a free parameter that we chose to match the observations). 
The resulting mixed tranet population reproduces well both the peaks and tails of the observed $\Delta$, period ratio, and $\xi$ distributions, in addition to the relative occurrence rates of single and multiple tranets in the observed multiplicity distribution. What remains to be explained is the population of planets with the smallest period ratios $\sim$1--1.5 and 
the tightest spacings $\Delta \lesssim 10$.
We speculate that these close neighbors may reflect orbital migration,
either of super-Earths or their progenitor cores.

Is a mixture of two populations truly necessary? The {\tt Ed$10^4$} ensemble is clearly a poor match to the observed distribution, but the improvement of the combined population over {\tt Ed$10^2$} is less obvious. 
We perform some statistical tests to help
decide this issue.
A two-sample K-S test applied to
the distribution of transit duration ratios (Fig.~14, row 4) leads us to reject the null hypothesis that the observed distribution and the {\tt Ed$10^2$} distribution are drawn from the same distribution ($p$-value 0.0004); at the same time, we cannot reject the null hypothesis that the observed and combined-group period ratios are drawn from the same distribution ($p$-value 0.02). Applying the K-S test to the
period ratio distributions (row 2), we reject the null hypothesis that
the observed period ratios
are drawn from any of the simulated
distributions -- if we consider the full
range of period ratios. But if we restrict
our attention to period ratios greater than
that of the 3:2 resonance, we find that we cannot reject the null hypothesis that the observed and combined-group period ratios are drawn from the same distribution ($p$-value 0.06), while still rejecting
the null hypothesis that the observed period ratios and {\tt Ed$10^2$} period ratios are drawn from the same distribution ($p$-value 0.004). 
When we apply the K-S test to the
$\Delta$ distributions (row 1),
we reject the null hypothesis that the
observed spacings are drawn from any of the simulated distributions ({\tt Ed$10^4$}, {\tt Ed$10^2$}, or combined) --- even if
we make a cut for $\Delta > 20$.
Finally, we consider the tranet multiplicity distribution (row 3), for which a K-S test is inappropriate because the distribution is spread across a small number of integers. If we assume Poisson uncertainties and add them in quadrature, the observed ratios of tranet multiplicities are $3.55 \pm 0.24, 2.78 \pm 0.33, 2.6 \pm 0.5, 3.5 \pm 1.2$, and $6 \pm 4$ for 1:2, 2:3, 3:4, 4:5, and 5:6 tranet systems, respectively. In {\tt Ed$10^2$}, these ratios are 2.5, 3.6, 1.7, 2.6, and 2.3 (chi-squared of 27), whereas for the mixture model, these ratios are 3.6, 4.3, 2.0, 2.7, and 2.3 (chi-squared of 23). Therefore the mixture model represents an improved  match to the observations; it matches the observed distribution well except that it overproduces two tranet systems relative to three tranet systems. 
In sum, the mixture model offers quantitative improvements for matching three out of the four observables in Fig.~14.

The purpose of the K-S and chi-squared statistics computed here is to quantify 
the significance of features we see in the plots, such as a claimed difference between two distributions, or a putative improvement of a mixture model over a single ensemble. Our aim is to ensure that the features discerned by eye are statistically significant and not due to chance and/or insufficient sample size. We caution that because of the simplifications in our simulations (including their limited 30 Myr timespan), these statistics should not be used to fine-tune our simulation parameters to match the observations.

 When we extended the simulated timescale of the {\tt Ed10$^2$} ensemble from 30 Myr to 180 Myr, the distributions of tranet spacing, period ratio, and duration ratio remained just as different from the observed distributions, but the distribution of tranet multiplicities changed to provide a better match to the observed distribution. As discussed in Section 5.2, performing extended timescale simulations is an important next step for a detailed quantitative comparison to the observations.

Even if a single one of our ensembles
could be made to adequately
match the observations,
our argument
that the data support two modes of planet
formation --- reflecting gas-rich
vs.~gas-poor environments --- would still
carry weight. 
 The ensemble {\tt Ed$10^2$}
is itself a mixture model to a degree,
since it contains simulations 
having a range of solid surface densities.
Core coagulation times are exponentially sensitive
to solid surface density \citep{Daws15}.
For solid surface densities toward
the upper end of the range in {\tt Ed$10^2$},
cores coagulate sufficiently quickly that gas is
still present at the end of coagulation.
For solid surface densities toward the
lower end of the range, cores coagulate more
slowly, in environments drained of gas.
The thesis of our paper is that the observations
implicate two modes of core coagulation for
super-Earths: a gas-rich mode (but
not too gas-rich; see \citealt{Lee14}
and \citealt{Lee16}) and a gas-poor mode.

A mixture of two populations is consistent with the statistical study by \citet{Ball14} of M-dwarf systems; using a parametric mixture model for the distribution of tranets, they found evidence for two populations, one spawning single tranets and another of high tranet multiplicity. \citet{Ball14} 
gave more weight to systems producing single tranets (1:1) than we do
(0.5:1). Here we find that a mixture reproduces not only the multiplicity statistics but also the $\xi$, $\Delta$, and period ratio distributions. In terms of the implications of our results for planet formation, our mixture represents a set of planets that underwent most of their growth in the presence of some residual gas ({\tt Ed$10^2$}), plus another set that assembled after the gas disk had completely dissipated ({\tt Ed$10^4$}; this ensemble could be replaced with one or more of the {\tt Ed$10^3$, Ed$10^1$, Ed$10^0$} ensembles with different weightings).

Among systems with two or more tranets, the median
eccentricity and inclination in ensemble {\tt Ed$10^2$} are 0.02 and 0.7$^\circ$, respectively. Without invoking tidal circularization, the median
eccentricity is in good agreement with the Rayleigh parameter $\sigma_e = 0.018 ^{+0.005}_{-0.004}$ (corresponding to a median $\tilde{e}=0.02$) measured by \citet{Hadd14} for planets with transit timing variations. Furthermore, among our simulated pairs
in {\tt Ed$10^2$} with period ratios less than 2,
the median eccentricity is as small
as $\tilde{e} = 0.003$. The ensemble {\tt Ed$10^2$} is the one that we argued in Section 6 should also produce the lowest density planets. In contrast, the ensembles that we argued should produce higher density planets have median tranet eccentricities of $\{0.11,0.08,0.08,0.11\}$ and inclinations of $\{1.6,1.6,0.9,1.9^\circ\}$ for ensembles {\tt$\{$Ed$10^0$, Ed$10^1$, Ed$10^3$, Ed$10^4\}$}, respectively. For period ratios less than 2, the corresponding median eccentricities are $\tilde{e} = \{0.07, 0.05, 0.05,$ $0.10\}$. The lower eccentricities and lower bulk densities of the {\tt Ed$10^2$} ensemble may account for the result of \citet{Hadd14} that larger planets ($\sim$2.5--4 $R_\oplus$) having lower densities \citep{Lope14,Roge15} have smaller eccentricities than planets $< 2.5 R_\oplus$. 

Another observational application of our simulations is to the finding of \citet{Weis14} that planets with masses measured via transit timing variations (TTVs, which necessarily
involve tight orbital spacings) have lower bulk densities than those measured via radial velocities
(which are easier to measure for more widely separated planets; see also \citealt{Wolf15}, and in particular \citealt{Stef15} for how selection effects based on mass may contribute to the result of \citealt{Weis14}). For the mixed ensemble plotted in the right panel of Fig.~\ref{fig:tranet}, period ratios less than 2 (i.e., period ratios typical for TTV planets) are dominated by planets in the {\tt Ed$10^2$} ensemble, which are expected to have lower densities. 
Furthermore, as described in Section 6.1, for a particular gas depletion factor and a range of solid surface densities, planets of a given mass having tighter spacings will also typically have lower densities.

\section{Conclusions}

\label{sec:conclude}

The circumstances of the giant impact era of planet formation link super-Earths' orbital properties to their compositions. Two types of systems are established: dynamically hot (widely spaced, eccentric, mutually inclined) planets with high bulk densities, and dynamically cold (tightly spaced, circular, flat) planets with gas envelopes and low densities.
Our Solar System's terrestrial planets fall into the former category. 
Typical orbital properties for the latter category --- Hill spacings $\Delta \sim 20$, $i \sim 0.8^\circ$, and $e \sim 0.02$ --- are dynamically colder.
The Hill spacing is set by an eccentricity equilibrium between gravitational scatterings, which tend to increase $e$, and mergers, which tend to damp it (Section 3).
This eccentricity equilibrium predicts that lower inclination systems
end up with smaller Hill spacings, a result that is confirmed numerically.

The two most important disk properties for determining the orbits and
compositions of planets are the disk's solid surface density and its
late-stage residual gas surface density (Section 6). Disks with
higher solid surface density and moderate gas surface density (depleted by a
factor of $\sim$100 relative to the minimum-mass solar nebula)
tend to produce tightly spaced, low $e$ and $i$, low density planets. 
Higher solid surface density enables cores to form faster --- fast enough
for cores to accrete volumetrically significant atmospheres from residual
gas before it dissipates completely
(Section 6.1; see also \citealt{Daws15} who make the same point).
A moderate late-stage gas surface density has enough dynamical friction to
flatten the system --- allowing for an
eccentricity equilibrium to be achieved
at tighter spacings (Section 3) --- but not so much that mergers
up to the final core mass are prevented. 
Higher gas surface densities prevent mergers,
so cores grow only after the disk gas dissipates sufficiently.
If gas surface densities
are too low --- and if the inner disk gas is 
not replenished (see \citealt{Lee16} for a discussion
of replenishment) --- then planets fail to acquire significant
atmospheres.
Hence a happy medium for the gas density is required
to produce tightly spaced, low density planets.

We found that we need a mixture of dynamically hot and dynamically cold
systems to reproduce the observed Hill spacings, period ratios, tranet
multiplicities, and transit duration ratios of \kep
super-Earths (Section 7). We can match the observed distributions with a
combination of moderate and lower (at the time of core assembly) gas surface densities,
in concert with a range of solid surface densities. However, we 
under-produce the tail of ultra tightly-spaced systems (period ratios $<$
1.5; e.g., Kepler-36, \citealt{Cart12}).
Our findings are consistent with previous work that modeled the multiplicity
distribution parametrically and found the need for two populations to match
the observed tranet multiplicity distribution \citep{Joha12,Ball14}. 
Without needing to invoke tidal circularization, our simulations with
moderate residual gas surface density can account for the low eccentricities
found by \citet{Hadd14} and \citet{Vane15}. Our finding that lower planetary
bulk densities are linked to tighter spacings and smaller eccentricities can
potentially account for the discovery by \citet{Hadd14} that larger
super-Earths (having lower densities, e.g., \citealt{Lope14,Roge15}) have
smaller eccentricities. It can also help to explain the result
of \citet{Weis14} that planets with masses measured via transit timing
variations (which are necessarily more tightly spaced) have
lower bulk densities.

Over the past couple decades, a variety of correlations have been uncovered
between the orbital and compositional properties of planetary bodies and properties of their host stars. These include the
connection between giant planet occurrence and host star metallicity (e.g.,
\citealt{Sant01,Fisc05}); the correlation between spin-orbit misalignment
and host star temperature (e.g., \citealt{Schl10,Winn10}); the relation
between planet radius and host star metallicity (e.g.,
\citealt{Buch14,Schla15,Daws15}); the correlation between
dynamical excitation and host star metallicity (e.g., \citealt{Daws13}); and the connection between Kuiper belt objects' compositional and orbital properties (e.g., \citealt{Tegl00,Levi01,Step06,Bruc09}).
The theoretical connections we found here between super-Earths' orbital
properties and their compositions provide a new test of formation theories.

The simplified simulations and first-order comparisons to observations
presented here can be improved upon. We caution that we did not incorporate
photo-evaporation of planetary atmospheres into our models so future studies that search for the
predicted links between orbital and compositional properties should either
account for photo-evaporation or be limited to periods beyond $\sim$15 days.
Eccentricity damping from residual gas could be incorporated explicitly into
the scaling arguments in Section 3. More realistic simulations could
implement prescriptions for atmospheric 
growth through accretion \citep{Lee14,Lee15} 
and collisional outcomes
\citep{Stew12,Schl15} into {\tt mercury6}. 
Longer integrations (e.g., to 10 Gyr instead of the 30 Myr reported here) are important for detailed quantitative comparisons to the observed sample, because the median spacing increases by about 10\% per decade in integration time (Section 5.2; see also \citealt{Pu15}).

Recently \citet{Mori15} showed that a combination of
planets formed from disks of solids with steep and shallow surface density
profiles (a range of $\alpha$; Eqn.~\ref{eqn:memb}) could reproduce the
observed distributions of multiplicity, period ratio, and transit duration
ratio. To distinguish between theories that rely on differences
in surface density slope (Moriarty \& Ballard)
or on differences in the normalizations of both gas and solid surface
densities (this work), future studies should test how tranet properties
vary with orbital period. They should also probe for correlations
between orbital and compositional properties --- correlations that
our work predicts.

A follow-up study focused on planets near resonance could determine
whether the gas damping treated here provides sufficient dissipation 
to produce the observed asymmetry in period ratios near the 3:2 and 2:1
resonances without invoking tides (e.g., \citealt{Lith12}).
Finally, the origin of the most tightly spaced planets (period ratios $< 1.5$)
remains a mystery and may be a signature of migration.
\vspace{0.2in}

\section*{Acknowledgments}
We thank the ApJ referee and the ApJ statistics editor for constructive and thought-provoking reports that led to substantive improvements in our paper. We also thank a MNRAS Letters referee for helpful comments that inspired us to expand this work into a more detailed and comprehensive study. We are grateful to Sarah Ballard, Daniel Fabrycky, Brad Hansen, Edwin Kite, Eiichiro Kokubo, Renu Malhotra, Ruth Murray-Clay, Hilke Schlichting, and Angie Wolfgang for informative discussions and comments. RID is supported by the Berkeley Miller Institute and EJL by NSERC under PGS D3 and the Berkeley Fellowship. EC thanks NSF
and NASA for financial support. The valuable collection of planet
candidates were discovered by NASA's \kep Mission and compiled from NASA's Exoplanet Archive, operated by Caltech, under contract with NASA under the Exoplanet Exploration Program. Simulations were run on the SAVIO computational cluster provided by Berkeley Research Computing.

%\bibliographystyle{apj}
%\bibliography{biblio}

\label{lastpage}

\end{document}